\documentclass{article}

\usepackage{graphicx}
\usepackage{amsmath}
\usepackage{subcaption}
\usepackage{etoc}
\usepackage{colortbl}
\usepackage{enumitem}

 \usepackage[main, final]{neurips_2026}

\usepackage{multirow}
\usepackage{pifont}

\usepackage[utf8]{inputenc} 
\usepackage[T1]{fontenc}    
\usepackage{hyperref}       
\usepackage{url}            
\usepackage{booktabs}       
\usepackage{amsfonts}       
\usepackage{nicefrac}       
\usepackage{microtype}      
\usepackage{xcolor}         
\usepackage{wrapfig}

\title{AffectGPT-RL: Revealing Roles of Reinforcement Learning in Open-Vocabulary Emotion Recognition}

\author{
Zheng Lian$^1$, 
Fan Zhang$^2$,
Lan Chen$^1$,
Yazhou Zhang$^3$,
Rui Liu$^4$,
Jinyang Wu$^5$, \\
\textbf{Haoyu Chen$^6$, 
Xiaobai Li$^7$,
Xiaojiang Peng$^8$,
Bin He$^1$,
Jianhua Tao$^5$}\\
\\[2mm]
$^1$State Key Laboratory of Autonomous Intelligent Unmanned Systems, Tongji University \\
$^2$The Chinese University of Hong Kong, 
$^3$Tianjin University, 
$^4$Inner Mongolia University \\
$^5$Tsinghua University, 
$^6$CMVS, University of Oulu, 
$^7$Zhejiang University \\
$^8$Shenzhen Technology University
}

\begin{document}

\maketitle

\begin{abstract}

Open-Vocabulary Multimodal Emotion Recognition (OV-MER) aims to predict emotions without being constrained by predefined label spaces, thereby enabling fine-grained emotion understanding. Unlike traditional discriminative methods, OV-MER leverages generative models to capture the full spectrum of emotions and employs emotion wheels (EWs) for metric calculation. Previous approaches primarily rely on token-level loss during training. However, this objective is misaligned with the metrics used in OV-MER, and these metrics cannot be directly optimized via gradient backpropagation. To address this limitation, we turn our attention to reinforcement learning, as this strategy can optimize non-differentiable objectives. We term this framework \textbf{AffectGPT-RL}. Furthermore, we conduct extensive experiments to elucidate the role of reinforcement learning in this task, revealing the necessity of the reasoning process, the impact of different rewards, and the generalizability to other emotion tasks such as sentiment analysis and basic emotion recognition. Experimental results demonstrate that AffectGPT-RL yields significant performance improvements on OV-MER. Beyond this task, we also achieve remarkable performance gains on basic emotion recognition, attaining state-of-the-art results on MER-UniBench. To the best of our knowledge, this is the pioneering work exploring the role of reinforcement learning in OV-MER, providing valuable guidance for subsequent researchers. Our code is provided at: \href{https://github.com/zeroQiaoba/AffectGPT/tree/master/AffectGPT-RL}{https://github.com/zeroQiaoba/AffectGPT/tree/master/AffectGPT-RL}.

\end{abstract}

\section{Introduction}
\label{sec:intro}

Emotion is highly related to human cognition, decision-making, and behavior, playing a vital role in our daily life \cite{huang2026grading,picard2000affective,xie2024emovit}. Naturally, humans express emotions through multiple modalities \cite{li2023multi}, including (micro)-gestures, (micro)-expressions, audio tones \cite{liu2024emotion,hu2025chain}, linguistic content \cite{li2025uni}, and other cues \cite{ben2021video,chen2023smg,el2011survey,lian2026merbench}. To better understand human emotions, it is essential to integrate these clues, enabling models to capture emotional nuances \cite{liu2025hardness}. This has led to the development of Multimodal Emotion Recognition (MER) \cite{lian2023mer,zhang2025moda,li2025uni}. Traditional MER approaches primarily rely on Ekman’s theory \cite{ekman1992argument,li2023decoupled,liu2024contrastive}, which categorizes human emotions into six basic labels: \emph{anger}, \emph{disgust}, \emph{happiness}, \emph{sadness}, \emph{fear}, and \emph{surprise}. However, human emotions extend far beyond these basic categories \cite{demszky2020goemotions,plutchik1980general}, and such a restricted label space inevitably leads to imprecise emotion representations.

\begin{wrapfigure}{r}{0.5\linewidth}
	\centering
	\includegraphics[width=\linewidth]{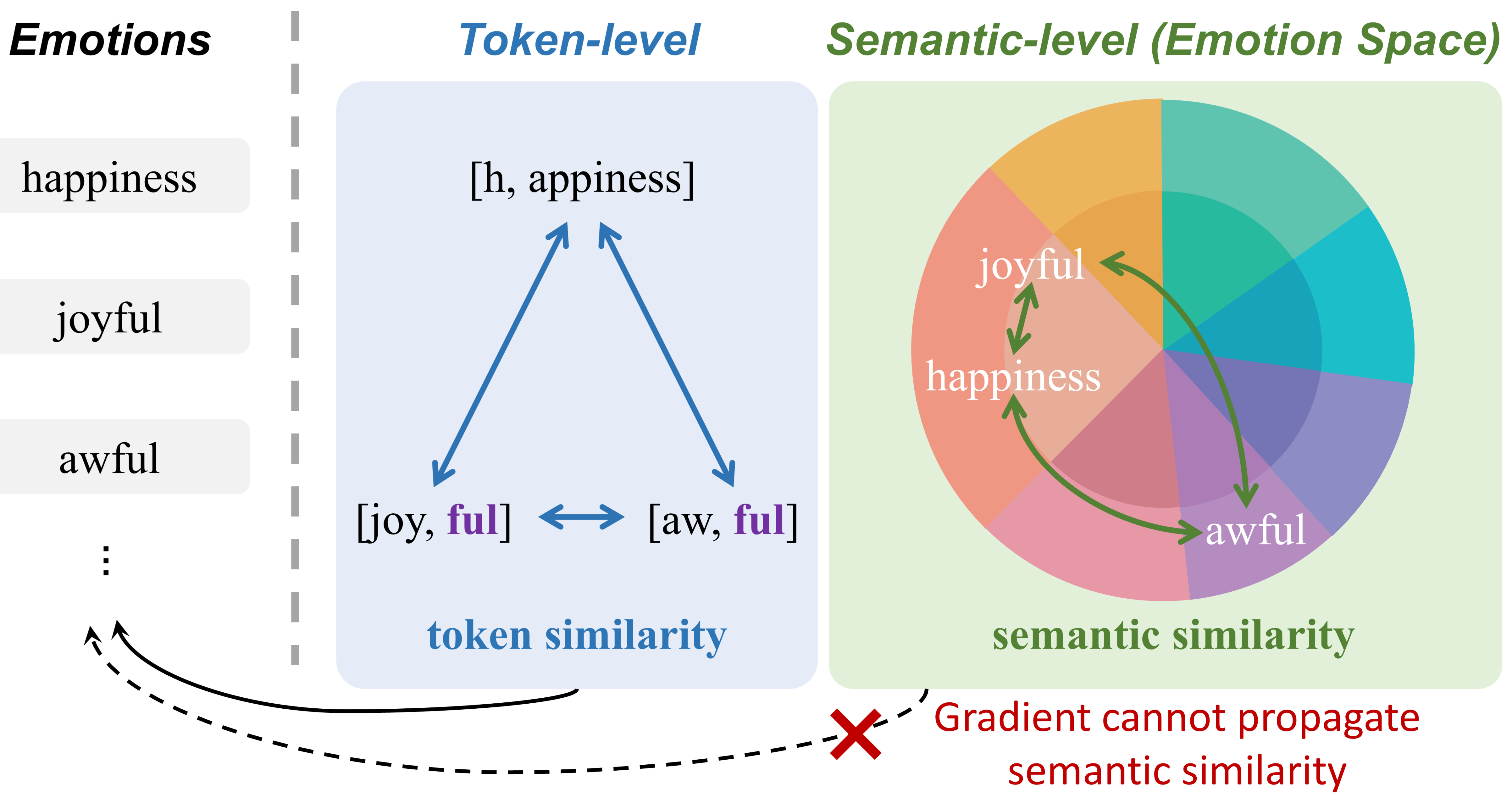}
	\caption{\textbf{Motivation.} Token similarity differs from semantic similarity. Existing work relies on token similarity, whereas the evaluation metrics for OV-MER depend on semantic similarity. Moreover, semantic similarity is inherently non-differentiable. This paper aims to address these issues through reinforcement learning.}
	\label{fig:overall}
    \vspace{-0.4cm}
\end{wrapfigure}

Recently, open-vocabulary MER (OV-MER) has emerged as a promising research direction, shifting emotion recognition from constrained categories to the full spectrum, thereby enabling more nuanced emotion representations \cite{lian2025affectgpt,lian2025ov}. To support this paradigm shift, OV-MER introduces new solutions and metrics. Regarding solutions, it transitions from discriminative to generative approaches, leveraging the extensive vocabulary of large language models (LLMs) to expand the recognition scope. Regarding metrics, it employs emotion wheels (EWs) to capture semantic relationships between distinct emotions. Appendix \ref{appendix:ew_calculate} provides a detailed illustration of these metrics. Current works primarily address this task by employing token-level loss to align predicted and ground-truth labels \cite{cheng2024emotion,lian2025affectgpt}. However, these methods suffer from a critical misalignment: token-level loss exhibits limited correlation with EW-based metrics. For instance, \emph{joyful} and \emph{awful} share high token-level similarity yet possess low semantic similarity in terms of their underlying emotions. Furthermore, since EW-based metrics are non-differentiable (see Appendix \ref{appendix:ew_nondiff}), they cannot be directly optimized via gradient backpropagation, posing significant challenges for existing solutions. Figure \ref{fig:overall} illustrates the motivation behind our work.
\vspace{-0.2cm}

To address these limitations, we shift our focus to reinforcement learning, as this strategy enables the optimization of non-differentiable objectives. We term this framework \textbf{AffectGPT-RL}. Specifically, we utilize EW-based metrics as reward functions and apply reinforcement learning to maximize these rewards, facilitating the direct optimization of these metrics. Meanwhile, we conduct extensive experiments to elucidate the role of reinforcement learning in OV-MER, including the necessity of the reasoning process, the influence of different reward functions, and the impact of reinforcement learning on other emotion tasks, such as sentiment analysis and basic emotion recognition. During training, we observe the issue of reward hacking \cite{everitt2017reinforcement}, where the model tends to generate lengthy predictions containing redundant emotional expressions. To mitigate this, we introduce length penalties into the reward function, encouraging more concise outputs while maintaining good performance. Experimental results demonstrate that AffectGPT-RL achieves significant improvements on OV-MER. Beyond this task, our framework also yields notable performance gains in basic emotion recognition, attaining state-of-the-art performance on MER-UniBench \cite{lian2025affectgpt}. \emph{This paper should not be viewed merely as an application of reinforcement learning, but rather as a significant milestone in advancing open-vocabulary emotion research. This work will inspire more researchers to consider reinforcement learning as a foundational tool for addressing this task.} The main contributions of this paper are summarized below:

\begin{itemize}[leftmargin=1.0em]

    \item \textbf{(Framework)} This is the first work to reveal the role of reinforcement learning in OV-MER.

    \item \textbf{(Experiment)} We conduct systematic experiments, revealing the role of the reasoning process, the impact of different rewards, and the generalization of reinforcement learning to other tasks.
    
    \item \textbf{(Performance)} Our method achieves state‑of‑the‑art performance on fine‑grained and basic emotion recognition, highlighting the potential of reinforcement learning in affective computing.
    
\end{itemize}

\section{AffectGPT-RL}
\label{sec:affectgpt_r1}
This paper aims to elucidate the role of reinforcement learning in OV-MER. Following previous works \cite{shao2024deepseekmath}, our training pipeline comprises two phases: cold-start training and reinforcement learning. In the first phase, we use a \emph{large-scale, coarse-grained} dataset to establish initial capabilities in emotion understanding and format alignment. In the second phase, we use a \emph{high-quality, fine-grained} dataset for reward calculation and policy updates. Figure \ref{fig:pipeline} illustrates the overall pipeline of AffectGPT-RL.

\subsection{Notation Definition}
\begin{figure*}[t]
	\centering
	\includegraphics[width=\linewidth]{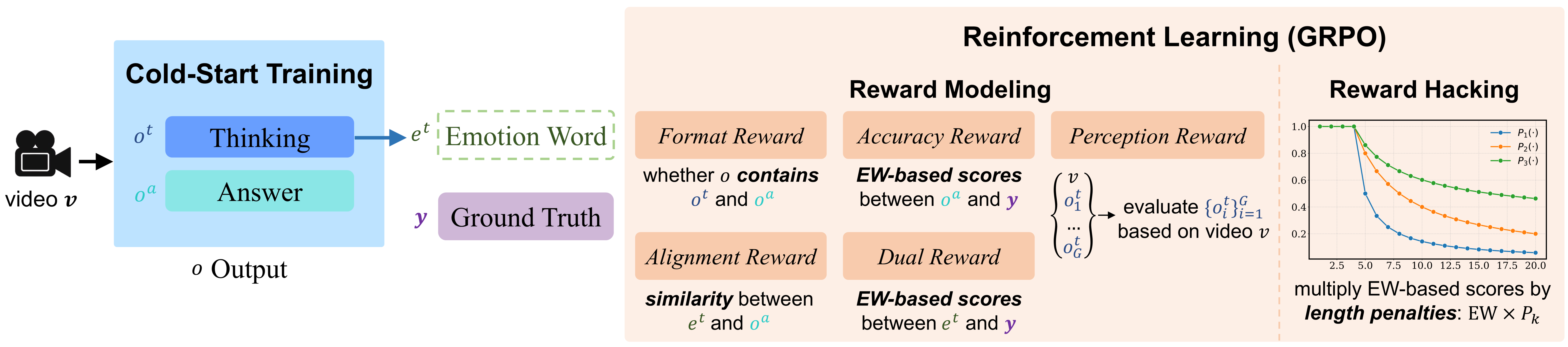}
	\caption{\textbf{AffectGPT-RL.} Our training process consists of two phases: cold-start training and reinforcement learning. In the first phase, we output both thinking and answers. In the second phase, we propose five rewards and introduce length penalties to mitigate reward hacking.}
	\label{fig:pipeline}
\end{figure*}

This section first clarifies the necessary notations. In Figure \ref{fig:pipeline}, we use $v$ to represent the video and $q$ to denote the user message. The output $o$ consists of two components: the thinking $o^t$ and the answer $o^a$. Specifically, $o^t$ and $o^a$ correspond to the content enclosed within the $<$think$>$ and $<$/think$>$ tags and the $<$answer$>$ and $<$/answer$>$ tags, respectively. We denote the ground-truth open-vocabulary labels as $y$. Our main goal is to predict the emotion $y$ based on the video content $v$. This paper focuses on utterance-level emotion recognition, ensuring fair comparison with prior work \cite{lian2025affectgpt,lian2025ov}. Appendix~\ref{appendix:task} provides additional clarification regarding our task formulation.

\subsection{Cold-Start Training}
Human emotions are highly dependent on subtle, multimodal cues \cite{li2022cas,li2022deep}, posing significant challenges for current models to capture all these signals and accurately interpret emotional states \cite{lian2024gpt,lu2024gpt}. To enhance emotion understanding, we employ {large-scale} descriptive emotion datasets for cold-start training. Specifically, we output both reasoning (enclosed within $<$think$>$ and $<$/think$>$ tags) and answers (enclosed within $<$answer$>$ and $<$/answer$>$ tags). Since \emph{descriptive} emotion datasets already follow a reasoning-like format \cite{lian2023explainable}, we treat these descriptions as $o^t$. Given our primary focus on OV-MER, we use \emph{open-vocabulary} emotions as $o^a$. This output format unifies \emph{descriptive} and \emph{open-vocabulary} emotions into a single task, enabling us to investigate whether these two tasks can mutually benefit each other. As demonstrated in Section \ref{sec:experiment_cold_start_data}, the quantity of cold-start data significantly affects both cold-start training performance and subsequent reinforcement learning, which further underscores our rationale for utilizing {large-scale} datasets in this phase.

\subsection{Reinforcement Learning}
\label{sec:reinforcement_learning}
Reinforcement learning relies on policy optimization and reward design. Specifically, we employ GRPO \cite{shao2024deepseekmath} for policy updates and propose five reward functions, all designed to guide the reasoning and answers. In this paper, we focus on designing reward models that better address OV-MER and leave the exploration of optimization methods to future work.

\subsubsection{Reward Modeling}
\paragraph{Format Reward.} 
To ensure structured outputs, we define a binary reward function that evaluates whether the model's response contains both thinking and answer components.
\begin{equation}
    \mathcal{R}_{\text{format}}(o|v, q, y) = \begin{cases}
        1, \mathrm{is\_required\_format}(o) \\
        0, \text{otherwise} \\
    \end{cases}
\end{equation}

\paragraph{Accuracy Reward.}
We employ the official metrics of OV-MER to compute \emph{accuracy rewards} \cite{lian2025ov}. Unlike discriminative MER, which centers on a fixed label space, OV-MER imposes no constraints on the label space. Consequently, models may generate synonyms, i.e., emotions with identical meanings but different expressions, which poses challenges for evaluation. To address this, OV-MER introduces EW-based metrics, which leverage the structural information of emotion wheels for clustering. Appendix~\ref{appendix:ew_metric} provides a detailed calculation procedure for these metrics.
\begin{equation}
    \mathcal{R}_{\text{accuracy}}(o|v, q, y) = \text{EW}\left(o^a, y\right).
\end{equation}

\paragraph{Alignment Reward.} \emph{Accuracy reward} ensures the correctness of $o^a$. However, without constraints on the reasoning process, the generated output may include irrelevant $o^a$ and $o^t$. To address this, we introduce \emph{alignment reward}, ensuring that the reasoning process is consistent with the final answer. As illustrated in Figure \ref{fig:pipeline}, the model output consists of both the thinking $o^t$ and the answer $o^a$. Specifically, we first extract emotion words $e^t$ from $o^t$. We then calculate the similarity between $e^t$ and $o^a$, serving as the alignment reward. By optimizing this reward, we encourage correlation between the reasoning process $o^t$ and the predicted answer $o^a$. Appendix \ref{appendix:alignment} provides the detailed methods for emotion word extraction and similarity calculation.
\begin{equation}
    \mathcal{R}_{\text{alignment}}(o|v, q,y) = \mathrm{is\_similar}\left(o^t \rightarrow e^t, o^a\right).
\end{equation}

\paragraph{Dual Reward.} Similar to \emph{alignment reward}, this reward also imposes constraints on $o^t$. Specifically, we require the emotion words extracted from $o^t$ to match the true label $y$:
\begin{equation}
    \mathcal{R}_{\text{dual}}(o|v, q, y) = \text{EW}\left(o^t \rightarrow e^t, y\right).
\end{equation}

\paragraph{Perception Reward.}
This reward constrains the thinking output $o^t$ to ensure its alignment with the video content. To achieve this, we draw inspiration from EmoPrefer \cite{lian2026emoprefer}, which proves that the MLLM can serve as a judge to compare two descriptions ($d_1$ and $d_2$) and determine which better reflects the emotional state of characters in a given video. To extend this to multiple descriptions, we propose to use the Bradley-Terry algorithm. Specifically, suppose we have $N$ descriptions $\{d_i\}_{i=1}^N$. We first compute all pairwise preferences using the MLLM, and then apply the Bradley-Terry algorithm to estimate the relative strength of each description. Finally, outputs ranked in the top 50\% receive a reward of 1, while the rest receive 0, thereby encouraging the generated reasoning to remain faithful to the video. Appendix \ref{appendix:perception} provides further details regarding the \emph{perception reward}.
\begin{equation}
    \mathcal{R}_{\text{perception}}(\{o_i\}_{i=1}^{G}|v,q,y) = \begin{cases}
        1, \mathrm{is\_top50}(o_i^t|v) \\
        0, \text{otherwise} \\
    \end{cases}
\end{equation}

\subsubsection{Reward Hacking}
\label{sec:method_reward_hacking}
During inference, we observe that the model tends to generate lengthy $o^a$ containing numerous emotional words. This is primarily because EW-based metrics automatically disregard synonyms during calculation (see Appendix \ref{appendix:ew_metric}). Consequently, synonyms do not influence the metric scores, whereas predicting more words increases the likelihood of matching the ground truth. This results in verbose outputs, which in turn increases inference time and reduces the readability of $o^a$. This phenomenon, also known as \emph{reward hacking} \cite{everitt2017reinforcement}, occurs when a model exploits loopholes or flaws in the reward function design to achieve high rewards while exhibiting undesired behavior. To address this, we introduce length penalties, encouraging the model to maximize rewards with fewer prediction words. Specifically, we propose three types of penalties as follows:
\begin{equation}
\begin{aligned}
P_{1} = 
\begin{cases}
1, & |o^a| \le |y| \\[2pt]
\frac{1}{|o^a| - |y| + 1}, & |o^a| > |y|
\end{cases}, \;
P_{2} = 
\begin{cases}
1, & |o^a| \le |y| \\[2pt]
\frac{|y|}{|o^a|}, & |o^a| > |y|
\end{cases}, \;
P_{3} = 
\begin{cases}
1, & |o^a| \le |y| \\[2pt]
\frac{\log|y|}{\log|o^a|}, & |o^a| > |y|
\end{cases}
\end{aligned}
\end{equation}
where $|o^a|$ and $|y|$ denote the number of emotion words in the predictions and ground truth, respectively. Figure \ref{fig:pipeline} visualizes the values of these penalties (see the \emph{reward hacking} part). From $P_{1}(\cdot)$ to $P_{3}(\cdot)$, the penalty on prediction length progressively loosens. We then apply these length penalties to the reward functions that include EW-based metrics: \emph{accuracy reward} and \emph{dual reward}.
\begin{align}
    &\mathcal{R}_{\text{accuracy}}(o|v, q, y) = P_k(o^a, y) \times \text{EW}(o^a, y),\\
    &\mathcal{R}_{\text{dual}}(o|v, q, y) = P_k(o^t \rightarrow e^t, y)\times\text{EW}(o^t \rightarrow e^t, y).
\end{align}



\section{Experimental Setup}
\label{sec:experiment_setup}

\paragraph{Corpus Description.}
AffectGPT-RL consists of two training phases. In the cold-start phase, we utilize a \emph{large-scale} dataset to endow the model with emotion understanding and format alignment capabilities. For this purpose, we select MER-Caption+ \cite{lian2025affectgpt}, which contains 31K samples annotated with rich emotion descriptions and open-vocabulary labels. In the reinforcement learning phase, we require a \emph{high-quality} dataset specifically designed for OV-MER. Accordingly, we choose MER2025-OV \cite{lian2025mer}, which features human-annotated, high-quality open-vocabulary labels. For evaluation, we assess the model’s performance on OV-MERD+ \cite{lian2025affectgpt} and MER-UniBench \cite{lian2025affectgpt} to measure its capabilities in open-vocabulary and other emotion tasks. We conduct a rigorous overlap check to ensure no data leakage exists between the training and testing sets (see Appendix~\ref{appendix:data_leakage}). Table \ref{tab:dataset} provides dataset statistics. During cold-start training, we leverage both descriptive and open-vocabulary emotions. However, in reinforcement learning \cite{wu2018learning}, we exclusively use open-vocabulary emotions. This decision stems from the fact that the reward calculation does not rely on human-annotated thinking results. Therefore, our dataset selection is well-suited to the distinct requirements of each training stage. For testing, besides open-vocabulary emotions, we also evaluate performance on other emotion tasks to reveal the generalization of reinforcement learning to these tasks.

\begin{table}[t]
	\centering
	\caption{\textbf{Dataset statistics.} This table provides statistics for the training and testing sets. We rigorously verify that no data leakage exists between these two sets (see Appendix~\ref{appendix:data_leakage}).}
    \label{tab:dataset}
    \resizebox{\linewidth}{!}{
        \begin{tabular}{l|r|cccc}
            \toprule
            \multirow{2}{*}{\textbf{Dataset}} & \multirow{2}{*}{\textbf{\#Samples}} & \multicolumn{4}{c}{\textbf{Emotion Recognition Task}} \\
            & & Sentiment Analysis & Basic Emotion & Open-Vocabulary Emotion & Descriptive Emotion \\
            \midrule
            \rowcolor{gray!20}
            \multicolumn{6}{c}{\emph{Training Phase 1: Cold-start training}} \\
            \midrule
            MER-Caption+ \cite{lian2025affectgpt} & 31,327 & $\times$ & $\times$ & $\surd$ & $\surd$ \\
            \midrule
            \rowcolor{gray!20}
            \multicolumn{6}{c}{\emph{Training Phase 2: Reinforcement learning}} \\
            \midrule
            MER2025-OV   \cite{lian2025mer}    & 1,000  & $\times$ & $\times$ & $\surd$ & $\times$ \\
            \midrule
            \rowcolor{gray!20}
            \multicolumn{6}{c}{\emph{Testing}} \\
            \midrule
            OV-MERD+     \cite{lian2025affectgpt}     & 532    & $\times$ & $\times$ & $\surd$ & $\times$ \\
            MER-UniBench \cite{lian2025affectgpt} & 12,799     & $\surd$ & $\surd$ & $\surd$ & $\times$ \\
            \bottomrule
        \end{tabular}
    }
\end{table}

\paragraph{Implementation Details.}
\label{sec:implementation_details}
During training, we set the learning rate to 1e-5 and the maximum number of epochs to 60. The entire implementation is conducted using PyTorch, and the code is executed on an NVIDIA A100 GPU. Due to memory constraints, we use a batch size of 3 for cold-start training and a batch size of 1 for reinforcement learning. For the model architecture, we adopt the default hyperparameters of AffectGPT \cite{lian2025affectgpt}. This setup ensures a fair comparison with the baselines, thereby isolating the performance gains attributable to our proposed framework.

\section{Results and Discussion}
\label{sec:results_and_discussion}
In this section, we conduct a series of experiments examining the impact of different rewards (Section~\ref{sec:experiment_reward}), the necessity of the thinking process (Section~\ref{sec:experiment_thinking}), the role of different length penalties (Section~\ref{sec:experiment_penalty}), generalization to other emotion tasks (Section~\ref{sec:experiment_main_results}), and various ablation studies regarding hyper-parameter sensitivity and data scaling in cold-start training and reinforcement learning (Section~\ref{sec:experiment_ablation}). Ultimately, we aim to answer the key question: \emph{Is reinforcement learning a promising approach for addressing open-vocabulary emotion recognition?}

\subsection{Impact of Different Rewards}
\label{sec:experiment_reward}
\begin{figure*}[!t]
    \centering
    \includegraphics[width=\linewidth]{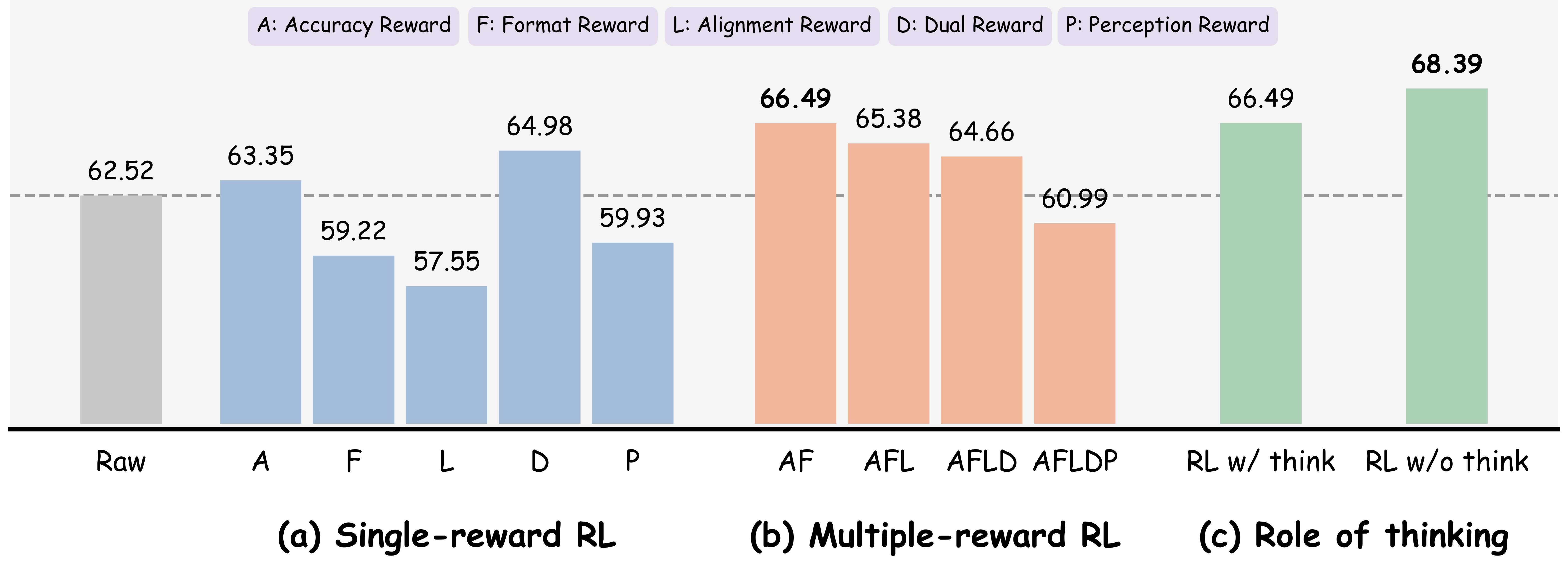}
    \caption{\emph{(a) Single-reward RL.} This figure reports performance obtained using RL with a single reward; \emph{(b) Multiple-reward RL.} It reports performance obtained using RL with multiple rewards; \emph{(c) Role of thinking.} It shows that a simpler model, one that only outputs the answer $o^a$ and calculates rewards based solely on $o^a$, achieves better performance.}
    \label{fig:merge}
\end{figure*}

This section evaluates the effectiveness of different reward functions. In Figure \ref{fig:merge}, we report performance under three scenarios: (1) \emph{without reinforcement learning}, (2) \emph{with reinforcement learning using a single reward}, and (3) \emph{with reinforcement learning using multiple rewards}.

\paragraph{Single-reward RL.}
In Figure \ref{fig:merge} (a), the \emph{accuracy reward} and \emph{dual reward} are the most effective, significantly improving performance compared to the model without reinforcement learning. These gains stem from the direct optimization of EW-based metrics via either the thinking $o^t$ or the answer $o^a$, highlighting the importance of aligning reinforcement learning objectives with evaluation metrics in OV-MER. In contrast, other rewards lead to performance degradation. For instance, the \emph{format reward} and \emph{alignment reward} show limited impact, as these aspects are already well-learned during the cold-start phase. The suboptimal performance of the \emph{perception reward} may be attributed to current MLLMs' inability to decode human preferences accurately, which can introduce reward noise \cite{lian2026emoprefer}. Interestingly, the \emph{accuracy reward} and \emph{dual reward} yield similar performance, suggesting that while EW-based metrics are crucial, the choice of optimizing $o^t$ or $o^a$ has minimal impact. Therefore, in subsequent experiments, we use the \emph{accuracy reward} by default.

\paragraph{Multiple-reward RL.}
In Figure \ref{fig:merge} (b), combining the \emph{accuracy reward} and \emph{format reward} yields further performance improvements over single-reward setups. However, employing more rewards, such as all five simultaneously, results in performance degradation. The primary issue stems from interference between rewards: optimizing for multiple objectives can dilute their individual effectiveness, ultimately compromising the outcome. For instance, emphasizing one reward may inadvertently diminish the influence of others. These findings suggest that more rewards do not necessarily equate to better performance. Instead, selecting an appropriate combination is crucial.

\subsection{Role of Thinking}
\label{sec:experiment_thinking}
In Section \ref{sec:experiment_reward}, the most effective rewards are the \emph{accuracy reward} and \emph{dual reward}, both of which are derived from evaluation metrics. This observation motivates a heuristic hypothesis: \emph{Could we modify the output mode to generate and optimize solely on the emotion words $o^a$, thereby encouraging the model to focus more on EW-based metrics?} Specifically, during cold-start training, we eliminate the thinking process and require the output to contain only $o^a$. During reinforcement learning, we optimize the \emph{accuracy reward} directly on $o^a$. We refer to the modified model as RL {(w/o think)} and the original model as RL {(w/ think)}. In Figure \ref{fig:merge} (c), removing the thinking process yields improved performance. This improvement can be attributed to two factors. First, the thinking process is derived from coarse-grained datasets, where inherent noise may lead to flawed reasoning and consequently degrade performance. Second, incorporating additional thinking content diverts the model’s attention from solving OV-MER, thereby increasing training difficulty. In future work, we plan to design high-quality cold-start data and conduct an in-depth analysis to determine for which samples or emotional states the thinking process is truly necessary.

\begin{wraptable}{r}{0.5\textwidth}
\centering
\caption{\textbf{Impact of different penalties.} Our goal is to identify a penalty that maintains model performance, prevents overlong outputs, and avoids introducing reward degradation. Based on these criteria, $P_2(\cdot)$ and $P_3(\cdot)$ are more suitable choices.}
\resizebox{\linewidth}{!}{
    \begin{tabular}{c|c|c|c}
        \toprule
        \textbf{Penalty} & \textbf{Performance} & \textbf{Overlong} & \textbf{Degradation} \\
        \midrule
        $P_1(\cdot)$  & 67.60  & \ding{55}  & \ding{52} \\
        $P_2(\cdot)$  & 67.79  & \ding{55}  & \ding{55}  \\
        $P_3(\cdot)$  & 68.05  & \ding{55}  & \ding{55}  \\
        \midrule
        None   & 68.39  & \ding{52} & \ding{55}  \\
        \bottomrule
    \end{tabular}
}
\label{tab:penalty}
\end{wraptable}
\subsection{Impact of Different Penalties}
\label{sec:experiment_penalty}
Following the discussion in Section \ref{sec:experiment_thinking}, we adopt RL (w/o think) as the default setting. During inference, we observe that the model tends to generate lengthy predictions containing redundant emotion words. This behavior stems from the suboptimal design of the EW-based metrics, where synonyms do not influence the metric scores, whereas predicting more words increases the likelihood of matching the ground truth (see Appendix \ref{appendix:ew_metric}). To mitigate this issue, we introduce three types of penalties in Section \ref{sec:method_reward_hacking}, ordered by increasing penalty strength from $P_3(\cdot)$ to $P_1(\cdot)$. This section evaluates the effectiveness of these penalties across three dimensions: (1) \emph{their impact on performance}; (2) \emph{their ability to prevent overlong outputs}; and (3) \emph{whether they introduce reward degradation}. As shown in Table \ref{tab:penalty}, all three penalties successfully address overlong outputs without compromising performance. However, the strictest penalty, $P_1(\cdot)$, leads to reward degradation. Therefore, $P_2(\cdot)$ and $P_3(\cdot)$ are preferable.

\subsection{Influence on Other Emotion Tasks}
\label{sec:experiment_main_results}
The preceding sections focus on open-vocabulary emotion recognition. In this section, we extend our evaluation to MER-UniBench \cite{lian2025affectgpt}, a unified benchmark comprising three key tasks: sentiment analysis, basic emotion recognition, and fine-grained emotion detection. Appendix~\ref{appendix:dataset} provides further details regarding this benchmark. This section aims to assess the generalization of reinforcement learning to other emotion tasks. To ensure a fair comparison, all methods are compared under the same input modality. Experimental results in Table~\ref{tab:main} demonstrate that AffectGPT-RL achieves significant performance improvements in both basic and fine-grained emotion recognition, but shows a slight decline in sentiment analysis. This decline represents an interesting negative result. The underlying cause is twofold. First, the model tends to over-generate nuanced emotional terms even when a simple \emph{positive} or \emph{negative} label would suffice, resulting in penalties under the EW-based metrics. Second, discrepancies exist among different tasks. OV-MER focuses on recognizing discrete emotion words (e.g., \emph{happy} and \emph{surprise}), which aligns closely with basic and fine-grained emotion recognition but diverges from the objectives of sentiment analysis. These factors collectively limit the effectiveness of reinforcement learning in sentiment analysis. Future work will focus on balancing reinforcement learning performance across different tasks. Overall, based on the average performance across the three tasks, AffectGPT-RL still achieves state-of-the-art results on MER-UniBench, validating the efficacy of reinforcement learning in affective computing.
\begin{table*}[h]
    \centering
	\renewcommand\arraystretch{1.1}
	\caption{\textbf{Influence on other emotion tasks.} MER-UniBench is a unified benchmark comprising three key tasks: sentiment analysis, basic emotion recognition, and fine-grained emotion detection. In this table, we aim to reveal the generalization of reinforcement learning to other emotion tasks.}
	\label{tab:main}
	\resizebox{\linewidth}{!}{
		\begin{tabular}{l|cccc|cccc|c|c}
			\toprule
                &\multicolumn{4}{c|}{\textbf{Sentiment}} 
                &\multicolumn{4}{c|}{\textbf{Basic}} 
                &{\textbf{Fine-grained}} &\multirow{2}{*}{\textbf{Mean}} \\
                &MOSI &MOSEI &SIMS &SIMS v2 & MER2023 & MER2024 & MELD &IEMOCAP & OV-MERD+ & \\
			
            \midrule
        \rowcolor{gray!20}
        \multicolumn{11}{c}{\emph{Input Modality: Audio, Text}} \\
        \midrule

OneLLM \cite{han2024onellm} & 64.01 & 54.09 & 63.39 & 61.98 & 25.52 & 17.21 & 28.32 & 33.44 & 22.25 & 41.14 \\
SECap \cite{xu2024secap} & 55.76 & 54.18 & 59.51 & 57.41 & 40.95 & 52.46 & 25.56 & 36.92 & 36.97 & 46.64 \\
PandaGPT \cite{su2023pandagpt} & 66.06 & 61.33 & 62.93 & 58.88 & 33.57 & 39.04 & 31.91 & 36.55 & 31.33 & 46.84 \\
Qwen-Audio \cite{chu2023qwen} & 70.09 & 46.90 & 70.73 & 65.26 & 41.85 & 31.61 & 49.09 & 35.47 & 32.36 & 49.26 \\
SALMONN \cite{tang2023salmonn} & 81.00 & 67.03 & 68.69 & 65.93 & 55.53 & 45.38 & 45.62 & 46.84 & 45.00 & 57.89 \\
AffectGPT \cite{lian2025affectgpt} & \textbf{83.46} & \textbf{80.74} & 82.99 & \textbf{83.75} & 72.94 & 73.41 & \textbf{56.63} & 55.68 & 59.98 & 72.18 \\
\textbf{AffectGPT-RL (Ours)} & 80.13 & 80.01 & \textbf{84.49} & 82.31 & \textbf{81.69} & \textbf{93.49} & \textbf{63.74} & \textbf{63.85} & \textbf{65.49} & \textbf{77.24} \\

        \midrule
        \rowcolor{gray!20}
        \multicolumn{11}{c}{\emph{Input Modality: Video, Text}} \\
        \midrule

Otter \cite{li2025otter} & 52.89 & 50.44 & 57.56 & 53.12 & 16.41 & 14.65 & 22.57 & 29.08 & 16.63 & 34.82 \\
Video-LLaVA \cite{lin2024video} & 56.37 & 61.64 & 53.28 & 57.45 & 36.93 & 30.25 & 30.73 & 38.95 & 34.00 & 44.40 \\
PandaGPT \cite{su2023pandagpt} & 58.50 & 64.25 & 62.07 & 65.25 & 39.13 & 47.16 & 38.33 & 47.21 & 35.07 & 50.77 \\
Video-ChatGPT \cite{maaz2024video} & 54.42 & 63.12 & 64.82 & 65.80 & 44.86 & 46.80 & 37.33 & 56.83 & 39.80 & 52.64 \\
VideoChat2 \cite{li2024mvbench} & 66.84 & 54.32 & 69.49 & 70.66 & 33.67 & 54.50 & 36.64 & 48.70 & 39.21 & 52.67 \\
LLaMA-VID \cite{li2024llama} & 61.78 & 63.89 & 69.35 & 67.48 & 50.72 & 57.60 & 42.75 & 46.02 & 45.01 & 56.07 \\
VideoChat \cite{li2025videochat} & 65.13 & 63.61 & 69.52 & 72.14 & 48.73 & 57.30 & 41.11 & 48.38 & 44.52 & 56.71 \\
Chat-UniVi \cite{jin2024chat} & 54.53 & 63.18 & 68.15 & 66.36 & 57.62 & 65.67 & 45.61 & 52.37 & 48.00 & 57.94 \\
mPLUG-Owl \cite{ye2023mplug} & 72.40 & 72.91 & 72.13 & 75.00 & 56.86 & 59.89 & 49.11 & 55.54 & 48.18 & 62.45 \\
AffectGPT \cite{lian2025affectgpt} & \textbf{82.39} & \textbf{81.57} & \textbf{87.20} & \textbf{86.29} & 74.58 & 75.29 & 57.63 & 62.19 & 61.65 & 74.31 \\
\textbf{AffectGPT-RL (Ours)} & 78.78 & 79.07 & 85.91 & 85.85 & \textbf{77.72} & \textbf{85.29} & \textbf{61.09} & \textbf{67.42} & \textbf{62.42} & \textbf{75.95} \\

        \midrule
        \rowcolor{gray!20}
        \multicolumn{11}{c}{\emph{Input Modality: Audio, Video, Text}} \\
        \midrule

PandaGPT \cite{su2023pandagpt} & 61.92 & 67.61 & 68.38 & 67.23 & 40.21 & 51.89 & 37.88 & 44.04 & 37.12 & 52.92 \\
R1-Omni \cite{zhao2025r1} & 58.02& 56.48& 71.82& 68.58& 64.17& 67.43& 43.20& 51.58& 55.24& 59.61 \\
Emotion-LLaMA \cite{cheng2024emotion} & 66.13 & 67.66 & 78.32 & 77.23 & 59.38 & 73.62 & 46.76 & 55.47 & 52.97 & 64.17 \\
AffectGPT \cite{lian2025affectgpt} & \textbf{81.30} & \textbf{80.90} & \textbf{88.49} & \textbf{86.18} & 78.54 & 78.80 & 55.65 & 60.54 & 62.52 & 74.77 \\
\textbf{AffectGPT-RL (Ours)} & 79.39 & 79.24 & 88.25 & 84.97 & \textbf{84.32} & \textbf{93.75} & \textbf{63.12} & \textbf{74.26} & \textbf{68.05} & \textbf{79.48} \\  

            \bottomrule
		\end{tabular}
	}
\end{table*}

\subsection{Ablation Study}
\label{sec:experiment_ablation}

\begin{figure*}[!t]
    \centering
    \includegraphics[width=\linewidth]{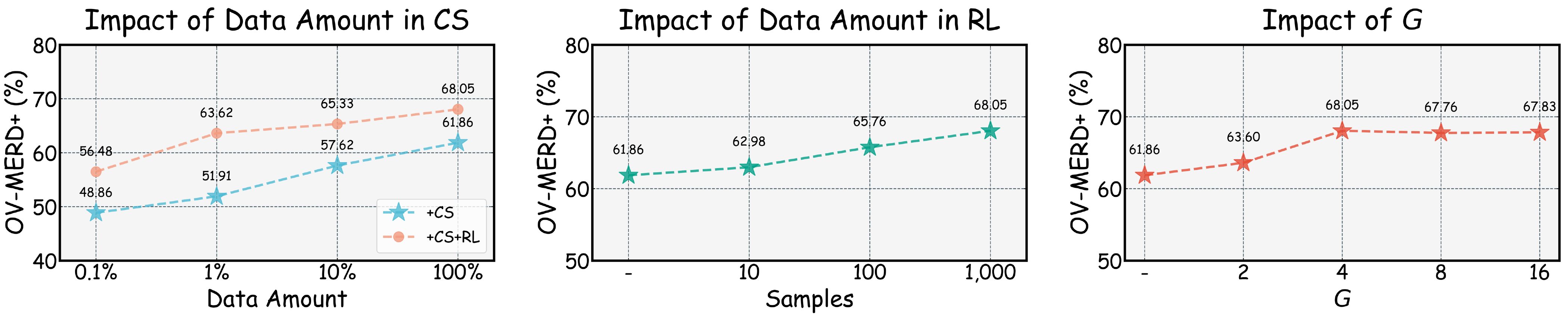}
    \caption{\textbf{Left: Impact of data amount in cold-start training.} This figure shows the performance of models trained with varying amounts of cold-start data and analyzes their influence on subsequent reinforcement learning. \textbf{Mid: Impact of data amount in reinforcement learning.} This figure shows that using a larger amount of data in reinforcement learning leads to better performance. \textbf{Right: Impact of $G$.} ``--'' denotes the baseline model without reinforcement learning.}
    \label{fig:ablation}
\end{figure*}

\paragraph{Data Amount in Cold-start Training.}
\label{sec:experiment_cold_start_data}

AffectGPT-RL consists of two stages: cold-start training (denoted as $s_1$) and reinforcement learning (denoted as $s_2$). In this section, we examine the impact of the amount of cold-start data on both $s_1$ and $s_2$. Experimental results shown in Figure \ref{fig:ablation} (left) indicate that increasing the amount of cold-start data improves performance in $s_1$. Furthermore, the performance after reinforcement learning strongly correlates with the performance of the base model from $s_1$, and additional cold-start data also benefits $s_2$. This is primarily because reinforcement learning mainly reinforces correct responses rather than introducing new knowledge \cite{liu2025understanding}. Therefore, a high-quality base model is crucial for maximizing the effectiveness of reinforcement learning.

\paragraph{Data Amount in Reinforcement Learning.}
\label{sec:experiment_data_rl}
This section investigates the impact of data quantity on reinforcement learning. Starting from the same model obtained via cold-start training (i.e., the model trained on 100\% of the data, as shown in the left part of Figure~\ref{fig:ablation}), we evaluate its performance using varying amounts of reinforcement learning data. Experimental results are presented in Figure~\ref{fig:ablation} (mid). We observe that utilizing just 10 samples for reinforcement learning yields performance improvements, confirming its effectiveness in addressing OV-MER. Furthermore, increasing the amount of training data leads to progressively greater performance gains. These trends suggest that AffectGPT-RL has not yet reached its performance upper bound. Therefore, future work will focus on collecting more data to further enhance the model's capabilities.

\paragraph{Impact of $G$.}
GRPO requires sampling $G$ outputs for group normalization. Figure \ref{fig:ablation} (right) reveals the impact of $G$. Due to GPU memory limitations, we restrict our comparisons to values of $G$ between 2 and 16 and do not evaluate larger values. Experimental results show that, regardless of the specific value of $G$, the model trained with reinforcement learning consistently outperforms the one without it, confirming the effectiveness of our approach. Meanwhile, the selection of $G$ also matters. A small $G$ does not fully exploit the benefits of GRPO, mainly because limited sampling reduces the chances of identifying optimal outcomes. In contrast, a larger $G$ (e.g., $G\geq4$) ensures the effectiveness of GRPO, but further increasing $G$ yields no significant improvement. Since a larger $G$ also increases computational cost, we set $G=4$ as the default configuration.

\paragraph{Stability Test.}
\label{sec:stability_test}
This section presents a stability test. Specifically, following the experiments in Section~\ref{sec:experiment_penalty}, we report the statistical results for single-run and multi-run settings. For the multi-run setting, we conduct the experiments three times and report both the mean score and the standard deviation. Experimental results are presented in Table~\ref{tab:penalty_stability}. From this table, we observe that AffectGPT-RL exhibits good stability across different runs. These results confirm that our conclusions are not artifacts of random chance, demonstrating the reliability of our experimental analysis.

\paragraph{Performance Gain Analysis.}
AffectGPT-RL consists of two training phases: cold-start training using MER-Caption+ and reinforcement learning using MER2025-OV. In Table \ref{tab:impact_of_cold_data}, we conduct a controlled experiment where AffectGPT is trained on the same two-stage dataset used by AffectGPT-RL. Experimental results show that even when using identical training data, the performance improvement of AffectGPT is marginal and significantly lags behind AffectGPT-RL. These results indicate that the performance gains stem from reinforcement learning rather than from the quantity of training data.

\begin{table}[t]
    \centering
    \label{tab:combined}
    \begin{minipage}[t]{0.38\linewidth}
    \centering
    \caption{\textbf{Stability test.} Following the experiments in Section~\ref{sec:experiment_penalty}, we report both single-run and multi-run results.}
    \resizebox{\linewidth}{!}{
        \begin{tabular}{c|c|c}
            \toprule
            \textbf{Penalty} & \textbf{Single-run} & \textbf{Multi-run} \\
            \midrule
            $P_1(\cdot)$ & 67.60 & 67.74 $\pm$ 0.15 \\
            $P_2(\cdot)$ & 67.79 & 68.05 $\pm$ 0.33 \\
            $P_3(\cdot)$ & 68.05 & 68.00 $\pm$ 0.10 \\
            \bottomrule
        \end{tabular}
    }
    \label{tab:penalty_stability}
    \\[0.5em]
    \end{minipage}
    \hfill
    \begin{minipage}[t]{0.59\linewidth}
    \centering
    \caption{\textbf{Performance gain analysis.} This table reveals that our performance gains stem from reinforcement learning rather than from an increase in data.}
    \resizebox{\linewidth}{!}{
        \begin{tabular}{l|l|c}
            \toprule
            \textbf{Model} & \textbf{Training Data} & \textbf{OV-MERD+} \\
            \midrule
            AffectGPT-RL & MER-Caption+ \& MER2025-OV  & 68.39 \\
            \midrule
            AffectGPT    & MER-Caption+                & 62.52 \\
            AffectGPT    & MER-Caption+ \& MER2025-OV  & 62.84 \\
            \bottomrule
        \end{tabular}
    }
    \label{tab:impact_of_cold_data}
    \\[0.5em]
    \end{minipage}
\end{table}

\section{Related Works}
\label{sec:formatting}

\subsection{Emotion Representation}

\paragraph{Dimensional Emotion Theory.} 
This theory uses continuous dimensions to capture the core attributes underlying emotions. For example, Russell \cite{russell1980circumplex} proposed a two-dimensional model, characterizing emotions based on \emph{valence} (the pleasantness of an emotion) and \emph{arousal} (the activation level of an emotion). In addition, some researchers have introduced a third dimension, \emph{dominance}, to represent the degree of perceived control over emotions \cite{russell1977evidence}. However, these dimensions are abstract and less intuitive for the general public, thus posing challenges in practical applications.

\paragraph{Basic Emotion Theory.} 
This theoretical framework typically relies on fixed and limited basic labels to describe the emotional states of characters. For example, Ekman \cite{ekman1992argument} argued that human emotions can be categorized into six basic labels, whereas Plutchik \cite{plutchik1980general} identified eight primary emotions that form the basis for other emotional experiences. Nevertheless, human emotions are rich and diverse, extending far beyond these basic labels. Mapping such nuanced emotional expressions onto a restricted set of basic labels inevitably leads to the loss of subtle emotional distinctions.

\paragraph{Open-Vocabulary Emotion.}
To address these limitations, researchers propose open-vocabulary representations \cite{lian2025ov}, which removes restrictions on the label space and enables models to predict any emotion. Unlike dimensional emotion theory, it focuses on more intuitive emotion words; Unlike basic emotion theory, it avoids restrictive label spaces, enabling more nuanced emotion representation. Therefore, the open-vocabulary paradigm opens up new opportunities for affective computing. However, this field is still in its early stages, and effective solutions remain unclear. This paper presents a pioneering effort that leverages reinforcement learning to tackle this task.

\subsection{Reinforcement Learning}

\paragraph{Policy Optimization.}
PPO \cite{schulman2017proximal} is a widely used algorithm that leverages a value model for advantage estimation and incorporates token-wise KL penalties to prevent excessive optimization. DPO \cite{rafailov2023direct} focuses on directly optimizing policies to align with human preferences, eliminating the need for an explicit reward model. GRPO \cite{shao2024deepseekmath} is a more efficient alternative to PPO, removing the value model and instead using a group-relative baseline computed from different outputs for the same input to guide optimization. Beyond these algorithms, some efforts have been made to enhance training efficiency and stability \cite{hu2025reinforce++, yu2025dapo}. This paper does not focus on designing new optimization algorithms, but rather on crafting optimal rewards for OV-MER.

\paragraph{Reward Modeling.}  
Reward systems can be broadly categorized into rule-based and model-based rewards. The former is suitable for tasks with verifiable outcomes. For example, pass or fail unit tests in code generation \cite{le2022coderl}, output language requirements in question-answering tasks \cite{guo2025deepseek}, and IoU between predicted bounding boxes and ground-truth annotations in object detection \cite{shen2025vlm}. The latter is mainly used for tasks where rewards are difficult to compute, such as safety, helpfulness, and harmlessness \cite{zhang2024safetybench}. In prior work, how to design reward functions tailored for OV-MER remains unexplored. To this end, we introduce various rewards to supervise both thinking and answers.

\section{Conclusion}
\label{sec:conclusion}
This paper proposes AffectGPT-RL, the first work to reveal the role of reinforcement learning in OV-MER. Through extensive experiments, we analyze the impact of different rewards and demonstrate the benefits of EW-based rewards. Meanwhile, we identify the limitations of incorporating a thinking process and show the usefulness of different-length penalties in addressing reward hacking. Beyond open-vocabulary emotions, we reveal the impact of reinforcement learning on other emotion tasks, demonstrating benefits for basic and fine-grained emotion recognition, while noting slight performance degradation in sentiment analysis. Furthermore, we investigate the effects of various modules and hyperparameters, confirming that our performance gains stem from reinforcement learning rather than other factors. These conclusions demonstrate the potential of reinforcement learning in open-vocabulary emotions and offer valuable insights for future research. This work will advance the development of this field toward a reinforcement learning era.

\section*{Acknowledgements}
This work is supported by the National Natural Science Foundation of China (62688201, 62671433, 62476146), Shanghai International Science and Technology Cooperation Project (26510713900), Open Research Fund of the State Key Laboratory of Multimodal Artificial Intelligence Systems (MAIS2026003), and the Innovative Team Project in Guangdong Province (2025KCXTD040).

\bibliographystyle{plain} 
\bibliography{mybib}

@article{shao2024deepseekmath,
  title={Deepseekmath: Pushing the limits of mathematical reasoning in open language models},
  author={Shao, Zhihong and Wang, Peiyi and Zhu, Qihao and Xu, Runxin and Song, Junxiao and Bi, Xiao and Zhang, Haowei and Zhang, Mingchuan and Li, YK and Wu, Yang and others},
  journal={arXiv preprint arXiv:2402.03300},
  year={2024}
}

@inproceedings{fang2025catch,
  title={Catch Your Emotion: Sharpening Emotion Perception in Multimodal Large Language Models},
  author={Fang, Yiyang and Liang, Jian and Huang, Wenke and Li, He and Su, Kehua and Ye, Mang},
  booktitle={Forty-second International Conference on Machine Learning},
  year={2025}
}

@article{zhao2025r1,
  title={R1-omni: Explainable omni-multimodal emotion recognition with reinforcement learning},
  author={Zhao, Jiaxing and Wei, Xihan and Bo, Liefeng},
  journal={arXiv preprint arXiv:2503.05379},
  year={2025}
}

@article{schulman2017proximal,
  title={Proximal policy optimization algorithms},
  author={Schulman, John and Wolski, Filip and Dhariwal, Prafulla and Radford, Alec and Klimov, Oleg},
  journal={arXiv preprint arXiv:1707.06347},
  year={2017}
}

@article{le2022coderl,
  title={Coderl: Mastering code generation through pretrained models and deep reinforcement learning},
  author={Le, Hung and Wang, Yue and Gotmare, Akhilesh Deepak and Savarese, Silvio and Hoi, Steven Chu Hong},
  journal={Advances in Neural Information Processing Systems},
  volume={35},
  pages={21314--21328},
  year={2022}
}

@article{rafailov2023direct,
  title={Direct preference optimization: Your language model is secretly a reward model},
  author={Rafailov, Rafael and Sharma, Archit and Mitchell, Eric and Manning, Christopher D and Ermon, Stefano and Finn, Chelsea},
  journal={Advances in Neural Information Processing Systems},
  volume={36},
  pages={53728--53741},
  year={2023}
}

@inproceedings{lian2025mer,
  title={Mer 2025: When affective computing meets large language models},
  author={Lian, Zheng and Liu, Rui and Xu, Kele and Liu, Bin and Liu, Xuefei and Zhang, Yazhou and Liu, Xin and Li, Yong and Cheng, Zebang and Zuo, Haolin and others},
  booktitle={Proceedings of the 33rd ACM International Conference on Multimedia},
  pages={13837--13842},
  year={2025}
}

@article{ekman1992argument,
  title={An argument for basic emotions},
  author={Ekman, Paul},
  journal={Cognition \& Emotion},
  volume={6},
  number={3-4},
  pages={169--200},
  year={1992},
  publisher={Taylor \& Francis}
}

@incollection{plutchik1980general,
  title={A general psychoevolutionary theory of emotion},
  author={Plutchik, Robert},
  booktitle={Theories of Emotion},
  pages={3--33},
  year={1980},
  publisher={Elsevier}
}

@inproceedings{lu2024gpt,
  title={Gpt as psychologist? preliminary evaluations for gpt-4v on visual affective computing},
  author={Lu, Hao and Niu, Xuesong and Wang, Jiyao and Wang, Yin and Hu, Qingyong and Tang, Jiaqi and Zhang, Yuting and Yuan, Kaishen and Huang, Bin and Yu, Zitong and others},
  booktitle={Proceedings of the IEEE/CVF Conference on Computer Vision and Pattern Recognition},
  pages={322--331},
  year={2024}
}

@inproceedings{xu2024secap,
  title={Secap: Speech emotion captioning with large language model},
  author={Xu, Yaoxun and Chen, Hangting and Yu, Jianwei and Huang, Qiaochu and Wu, Zhiyong and Zhang, Shi-Xiong and Li, Guangzhi and Luo, Yi and Gu, Rongzhi},
  booktitle={Proceedings of the AAAI Conference on Artificial Intelligence},
  pages={19323--19331},
  year={2024}
}

@inproceedings{li2024llama,
  title={Llama-vid: An image is worth 2 tokens in large language models},
  author={Li, Yanwei and Wang, Chengyao and Jia, Jiaya},
  booktitle={European Conference on Computer Vision},
  pages={323--340},
  year={2024},
  organization={Springer}
}

@inproceedings{jin2024chat,
  title={Chat-univi: Unified visual representation empowers large language models with image and video understanding},
  author={Jin, Peng and Takanobu, Ryuichi and Zhang, Wancai and Cao, Xiaochun and Yuan, Li},
  booktitle={Proceedings of the IEEE/CVF Conference on Computer Vision and Pattern Recognition},
  pages={13700--13710},
  year={2024}
}

@inproceedings{han2024onellm,
  title={Onellm: One framework to align all modalities with language},
  author={Han, Jiaming and Gong, Kaixiong and Zhang, Yiyuan and Wang, Jiaqi and Zhang, Kaipeng and Lin, Dahua and Qiao, Yu and Gao, Peng and Yue, Xiangyu},
  booktitle={Proceedings of the IEEE/CVF Conference on Computer Vision and Pattern Recognition},
  pages={26584--26595},
  year={2024}
}

@inproceedings{lin2024video,
  title={Video-LLaVA: Learning United Visual Representation by Alignment Before Projection},
  author={Lin, Bin and Ye, Yang and Zhu, Bin and Cui, Jiaxi and Ning, Munan and Jin, Peng and Yuan, Li},
  booktitle={Proceedings of the 2024 Conference on Empirical Methods in Natural Language Processing},
  pages={5971--5984},
  year={2024}
}

@article{lian2024gpt,
  title={Gpt-4v with emotion: A zero-shot benchmark for generalized emotion recognition},
  author={Lian, Zheng and Sun, Licai and Sun, Haiyang and Chen, Kang and Wen, Zhuofan and Gu, Hao and Liu, Bin and Tao, Jianhua},
  journal={Information Fusion},
  volume={108},
  pages={102367},
  year={2024},
  publisher={Elsevier}
}

@article{el2011survey,
  title={Survey on speech emotion recognition: Features, classification schemes, and databases},
  author={El Ayadi, Moataz and Kamel, Mohamed S and Karray, Fakhri},
  journal={Pattern Recognition},
  volume={44},
  number={3},
  pages={572--587},
  year={2011},
  publisher={Elsevier}
}

@inproceedings{demszky2020goemotions,
  title={GoEmotions: A Dataset of Fine-Grained Emotions},
  author={Demszky, Dorottya and Movshovitz-Attias, Dana and Ko, Jeongwoo and Cowen, Alan and Nemade, Gaurav and Ravi, Sujith},
  booktitle={Proceedings of the 58th Annual Meeting of the Association for Computational Linguistics},
  pages={4040--4054},
  year={2020}
}

@article{ben2021video,
  title={Video-based facial micro-expression analysis: A survey of datasets, features and algorithms},
  author={Ben, Xianye and Ren, Yi and Zhang, Junping and Wang, Su-Jing and Kpalma, Kidiyo and Meng, Weixiao and Liu, Yong-Jin},
  journal={IEEE Transactions on Pattern Analysis and Machine Intelligence},
  volume={44},
  number={9},
  pages={5826--5846},
  year={2021},
  publisher={IEEE}
}

@inproceedings{lian2024mer,
  title={Mer 2024: Semi-supervised learning, noise robustness, and open-vocabulary multimodal emotion recognition},
  author={Lian, Zheng and Sun, Haiyang and Sun, Licai and Wen, Zhuofan and Zhang, Siyuan and Chen, Shun and Gu, Hao and Zhao, Jinming and Ma, Ziyang and Chen, Xie and others},
  booktitle={Proceedings of the 2nd International Workshop on Multimodal and Responsible Affective Computing},
  pages={41--48},
  year={2024}
}

@article{chen2023smg,
  title={Smg: A micro-gesture dataset towards spontaneous body gestures for emotional stress state analysis},
  author={Chen, Haoyu and Shi, Henglin and Liu, Xin and Li, Xiaobai and Zhao, Guoying},
  journal={International Journal of Computer Vision},
  volume={131},
  number={6},
  pages={1346--1366},
  year={2023},
  publisher={Springer}
}

@inproceedings{zadeh2018multimodal,
  title={Multimodal language analysis in the wild: Cmu-mosei dataset and interpretable dynamic fusion graph},
  author={Zadeh, AmirAli Bagher and Liang, Paul Pu and Poria, Soujanya and Cambria, Erik and Morency, Louis-Philippe},
  booktitle={Proceedings of the 56th Annual Meeting of the Association for Computational Linguistics (Volume 1: Long Papers)},
  pages={2236--2246},
  year={2018}
}

@book{picard2000affective,
  title={Affective computing},
  author={Picard, Rosalind W},
  year={2000},
  publisher={MIT press}
}

@inproceedings{xie2024emovit,
  title={Emovit: Revolutionizing emotion insights with visual instruction tuning},
  author={Xie, Hongxia and Peng, Chu-Jun and Tseng, Yu-Wen and Chen, Hung-Jen and Hsu, Chan-Feng and Shuai, Hong-Han and Cheng, Wen-Huang},
  booktitle={Proceedings of the IEEE/CVF Conference on Computer Vision and Pattern Recognition},
  pages={26596--26605},
  year={2024}
}

@inproceedings{poria2019meld,
  title={Meld: A multimodal multi-party dataset for emotion recognition in conversations},
  author={Poria, Soujanya and Hazarika, Devamanyu and Majumder, Navonil and Naik, Gautam and Cambria, Erik and Mihalcea, Rada},
  booktitle={Proceedings of the 57th Conference of the Association for Computational Linguistics},
  pages={527--536},
  year={2019}
}

@inproceedings{zadeh2017tensor,
  title={Tensor fusion network for multimodal sentiment analysis},
  author={Zadeh, Amir and Chen, Minghai and Poria, Soujanya and Cambria, Erik and Morency, Louis-Philippe},
  booktitle={Proceedings of the Conference on Empirical Methods in Natural Language Processing},
  pages={1103--1114},
  year={2017}
}

@article{russell1980circumplex,
  title={A circumplex model of affect.},
  author={Russell, James A},
  journal={Journal of Personality and Social Psychology},
  volume={39},
  number={6},
  pages={1161},
  year={1980},
  publisher={American Psychological Association}
}

@article{russell1977evidence,
  title={Evidence for a three-factor theory of emotions},
  author={Russell, James A and Mehrabian, Albert},
  journal={Journal of Research in Personality},
  volume={11},
  number={3},
  pages={273--294},
  year={1977},
  publisher={Elsevier}
}

@article{guo2025deepseek,
  title={Deepseek-r1 incentivizes reasoning in llms through reinforcement learning},
  author={Guo, Daya and Yang, Dejian and Zhang, Haowei and Song, Junxiao and Wang, Peiyi and Zhu, Qihao and Xu, Runxin and Zhang, Ruoyu and Ma, Shirong and Bi, Xiao and others},
  journal={Nature},
  volume={645},
  number={8081},
  pages={633--638},
  year={2025},
  publisher={Nature Publishing Group UK London}
}

@article{liu2025understanding,
  title={Understanding r1-zero-like training: A critical perspective},
  author={Liu, Zichen and Chen, Changyu and Li, Wenjun and Qi, Penghui and Pang, Tianyu and Du, Chao and Lee, Wee Sun and Lin, Min},
  journal={arXiv preprint arXiv:2503.20783},
  year={2025}
}

@article{hu2025reinforce++,
  title={Reinforce++: A simple and efficient approach for aligning large language models},
  author={Hu, Jian},
  journal={arXiv preprint arXiv:2501.03262},
  year={2025}
}

@article{li2022deep,
  title={Deep learning for micro-expression recognition: A survey},
  author={Li, Yante and Wei, Jinsheng and Liu, Yang and Kauttonen, Janne and Zhao, Guoying},
  journal={IEEE Transactions on Affective Computing},
  volume={13},
  number={4},
  pages={2028--2046},
  year={2022},
  publisher={IEEE}
}

@article{li2022cas,
  title={CAS (ME) 3: A third generation facial spontaneous micro-expression database with depth information and high ecological validity},
  author={Li, Jingting and Dong, Zizhao and Lu, Shaoyuan and Wang, Su-Jing and Yan, Wen-Jing and Ma, Yinhuan and Liu, Ye and Huang, Changbing and Fu, Xiaolan},
  journal={IEEE Transactions on Pattern Analysis and Machine Intelligence},
  volume={45},
  number={3},
  pages={2782--2800},
  year={2022},
  publisher={IEEE}
}

@inproceedings{lian2026emoprefer,
  title={EmoPrefer: Can Large Language Models Understand Human Emotion Preferences?},
  author={Lian, Zheng and Sun, Licai and Chen, Lan and Chen, Haoyu and Cheng, Zebang and Zhang, Fan and Jia, Ziyu and Ma, Ziyang and Ma, Fei and Peng, Xiaojiang and others},
  booktitle={International Conference on Learning Representations},
  year={2026}
}

@inproceedings{yu2025dapo,
  title={Dapo: An open-source llm reinforcement learning system at scale},
  author={Yu, Qiying and Zhang, Zheng and Zhu, Ruofei and Yuan, Yufeng and Zuo, Xiaochen and Yue, Yu and Dai, Weinan and Fan, Tiantian and Liu, Gaohong and Liu, Lingjun and others},
  booktitle={Proceedings of the Advances in Neural Information Processing Systems},
  year={2025}
}

@inproceedings{zhang2024safetybench,
  title={SafetyBench: Evaluating the Safety of Large Language Models},
  author={Zhang, Zhexin and Lei, Leqi and Wu, Lindong and Sun, Rui and Huang, Yongkang and Long, Chong and Liu, Xiao and Lei, Xuanyu and Tang, Jie and Huang, Minlie},
  booktitle={Proceedings of the 62nd Annual Meeting of the Association for Computational Linguistics (Volume 1: Long Papers)},
  pages={15537--15553},
  year={2024}
}

@inproceedings{everitt2017reinforcement,
  title={Reinforcement learning with a corrupted reward channel},
  author={Everitt, Tom and Krakovna, Victoria and Orseau, Laurent and Legg, Shane},
  booktitle={Proceedings of the 26th International Joint Conference on Artificial Intelligence},
  pages={4705--4713},
  year={2017}
}

@article{shen2025vlm,
  title={Vlm-r1: A stable and generalizable r1-style large vision-language model},
  author={Shen, Haozhan and Liu, Peng and Li, Jingcheng and Fang, Chunxin and Ma, Yibo and Liao, Jiajia and Shen, Qiaoli and Zhang, Zilun and Zhao, Kangjia and Zhang, Qianqian and others},
  journal={arXiv preprint arXiv:2504.07615},
  year={2025}
}

@article{busso2008iemocap,
  title={IEMOCAP: Interactive emotional dyadic motion capture database},
  author={Busso, Carlos and Bulut, Murtaza and Lee, Chi-Chun and Kazemzadeh, Abe and Mower, Emily and Kim, Samuel and Chang, Jeannette N and Lee, Sungbok and Narayanan, Shrikanth S},
  journal={Language Resources and Evaluation},
  volume={42},
  pages={335--359},
  year={2008},
  publisher={Springer}
}

@inproceedings{liu2022make,
  title={Make acoustic and visual cues matter: Ch-sims v2. 0 dataset and av-mixup consistent module},
  author={Liu, Yihe and Yuan, Ziqi and Mao, Huisheng and Liang, Zhiyun and Yang, Wanqiuyue and Qiu, Yuanzhe and Cheng, Tie and Li, Xiaoteng and Xu, Hua and Gao, Kai},
  booktitle={Proceedings of the International Conference on Multimodal Interaction},
  pages={247--258},
  year={2022}
}

@inproceedings{lian2023mer,
  title={Mer 2023: Multi-label learning, modality robustness, and semi-supervised learning},
  author={Lian, Zheng and Sun, Haiyang and Sun, Licai and Chen, Kang and Xu, Mngyu and Wang, Kexin and Xu, Ke and He, Yu and Li, Ying and Zhao, Jinming and others},
  booktitle={Proceedings of the 31st ACM International Conference on Multimedia},
  pages={9610--9614},
  year={2023}
}

@article{li2025videochat,
  title={Videochat: Chat-centric video understanding},
  author={Li, KunChang and He, Yinan and Wang, Yi and Li, Yizhuo and Wang, Wenhai and Luo, Ping and Wang, Yali and Wang, Limin and Qiao, Yu},
  journal={Science China Information Sciences},
  volume={68},
  number={10},
  pages={200102},
  year={2025},
  publisher={Springer}
}

@inproceedings{su2023pandagpt,
  title={PandaGPT: One Model To Instruction-Follow Them All},
  author={Su, Yixuan and Lan, Tian and Li, Huayang and Xu, Jialu and Wang, Yan and Cai, Deng},
  booktitle={Proceedings of the 1st Workshop on Taming Large Language Models: Controllability in the era of Interactive Assistants},
  pages={11--23},
  year={2023}
}

@inproceedings{maaz2024video,
  title={Video-ChatGPT: Towards Detailed Video Understanding via Large Vision and Language Models},
  author={Maaz, Muhammad and Rasheed, Hanoona and Khan, Salman and Khan, Fahad},
  booktitle={Proceedings of the 62nd Annual Meeting of the Association for Computational Linguistics (Volume 1: Long Papers)},
  pages={12585--12602},
  year={2024}
}

@inproceedings{girdhar2023imagebind,
  title={Imagebind: One embedding space to bind them all},
  author={Girdhar, Rohit and El-Nouby, Alaaeldin and Liu, Zhuang and Singh, Mannat and Alwala, Kalyan Vasudev and Joulin, Armand and Misra, Ishan},
  booktitle={Proceedings of the IEEE/CVF Conference on Computer Vision and Pattern Recognition},
  pages={15180--15190},
  year={2023}
}

@inproceedings{li2024mvbench,
  title={MVBench: A Comprehensive Multi-modal Video Understanding Benchmark},
  author={Kunchang Li and Yali Wang and Yinan He and Yizhuo Li and Yi Wang and Yi Liu and Zun Wang and Jilan Xu and Guo Chen and Ping Luo and Limin Wang and Yu Qiao},
  booktitle={Proceedings of the IEEE/CVF Conference on Computer Vision and Pattern Recognition},
  year={2024}
}

@inproceedings{tang2023salmonn,
  title={SALMONN: Towards Generic Hearing Abilities for Large Language Models},
  author={Tang, Changli and Yu, Wenyi and Sun, Guangzhi and Chen, Xianzhao and Tan, Tian and Li, Wei and Lu, Lu and MA, Zejun and Zhang, Chao},
  booktitle={International Conference on Learning Representations},
  year={2023}
}

@article{ye2023mplug,
  title={mPLUG-Owl: Modularization Empowers Large Language Models with Multimodality},
  author={Ye, Qinghao and Xu, Haiyang and Xu, Guohai and Ye, Jiabo and Yan, Ming and Zhou, Yiyang and Wang, Junyang and Hu, Anwen and Shi, Pengcheng and Shi, Yaya and others},
  journal={arXiv preprint arXiv:2304.14178},
  year={2023}
}

@article{cheng2024emotion,
  title={Emotion-llama: Multimodal emotion recognition and reasoning with instruction tuning},
  author={Cheng, Zebang and Cheng, Zhi-Qi and He, Jun-Yan and Wang, Kai and Lin, Yuxiang and Lian, Zheng and Peng, Xiaojiang and Hauptmann, Alexander},
  journal={Advances in Neural Information Processing Systems},
  volume={37},
  pages={110805--110853},
  year={2024}
}

@inproceedings{yu2020ch,
  title={Ch-sims: A chinese multimodal sentiment analysis dataset with fine-grained annotation of modality},
  author={Yu, Wenmeng and Xu, Hua and Meng, Fanyang and Zhu, Yilin and Ma, Yixiao and Wu, Jiele and Zou, Jiyun and Yang, Kaicheng},
  booktitle={Proceedings of the 58th Annual Meeting of the Association for Computational Linguistics},
  pages={3718--3727},
  year={2020}
}

@article{awadalla2023openflamingo,
  title={Openflamingo: An open-source framework for training large autoregressive vision-language models},
  author={Awadalla, Anas and Gao, Irena and Gardner, Josh and Hessel, Jack and Hanafy, Yusuf and Zhu, Wanrong and Marathe, Kalyani and Bitton, Yonatan and Gadre, Samir and Sagawa, Shiori and others},
  journal={arXiv preprint arXiv:2308.01390},
  year={2023}
}

@inproceedings{chen2023beats,
  title={BEATs: audio pre-training with acoustic tokenizers},
  author={Chen, Sanyuan and Wu, Yu and Wang, Chengyi and Liu, Shujie and Tompkins, Daniel and Chen, Zhuo and Che, Wanxiang and Yu, Xiangzhan and Wei, Furu},
  booktitle={Proceedings of the 40th International Conference on Machine Learning},
  pages={5178--5193},
  year={2023}
}

@inproceedings{radford2023robust,
  title={Robust speech recognition via large-scale weak supervision},
  author={Radford, Alec and Kim, Jong Wook and Xu, Tao and Brockman, Greg and McLeavey, Christine and Sutskever, Ilya},
  booktitle={International Conference on Machine Learning},
  pages={28492--28518},
  year={2023},
  organization={PMLR}
}

@article{li2025otter,
  title={Otter: A multi-modal model with in-context instruction tuning},
  author={Li, Bo and Zhang, Yuanhan and Chen, Liangyu and Wang, Jinghao and Pu, Fanyi and Cahyono, Joshua Adrian and Yang, Jingkang and Li, Chunyuan and Liu, Ziwei},
  journal={IEEE Transactions on Pattern Analysis and Machine Intelligence},
  year={2025},
  publisher={IEEE}
}

@article{chu2023qwen,
  title={Qwen-audio: Advancing universal audio understanding via unified large-scale audio-language models},
  author={Chu, Yunfei and Xu, Jin and Zhou, Xiaohuan and Yang, Qian and Zhang, Shiliang and Yan, Zhijie and Zhou, Chang and Zhou, Jingren},
  journal={arXiv preprint arXiv:2311.07919},
  year={2023}
}

@article{lian2026merbench,
  title={Merbench: A unified evaluation benchmark for multimodal emotion recognition},
  author={Lian, Zheng and Sun, Licai and Ren, Yong and Gu, Hao and Sun, Haiyang and Chen, Lan and Liu, Bin and Tao, Jianhua},
  journal={IEEE Transactions on Pattern Analysis and Machine Intelligence},
  year={2026},
  publisher={IEEE}
}

@article{poria2019emotion,
  title={Emotion recognition in conversation: Research challenges, datasets, and recent advances},
  author={Poria, Soujanya and Majumder, Navonil and Mihalcea, Rada and Hovy, Eduard},
  journal={IEEE Access},
  volume={7},
  pages={100943--100953},
  year={2019},
  publisher={IEEE}
}

@article{sun2024svfap,
  title={Svfap: Self-supervised video facial affect perceiver},
  author={Sun, Licai and Lian, Zheng and Wang, Kexin and He, Yu and Xu, Mingyu and Sun, Haiyang and Liu, Bin and Tao, Jianhua},
  journal={IEEE Transactions on Affective Computing},
  year={2024},
  publisher={IEEE}
}

@article{lian2023explainable,
  title={Explainable multimodal emotion reasoning},
  author={Lian, Zheng and Sun, Licai and Xu, Mingyu and Sun, Haiyang and Xu, Ke and Wen, Zhuofan and Chen, Shun and Liu, Bin and Tao, Jianhua},
  journal={arXiv preprint arXiv:2306.15401},
  year={2023}
}

@inproceedings{lian2025ov,
  title={OV-MER: Towards Open-Vocabulary Multimodal Emotion Recognition},
  author={Lian, Zheng and Sun, Haiyang and Sun, Licai and Chen, Haoyu and Chen, Lan and Gu, Hao and Wen, Zhuofan and Chen, Shun and Siyuan, Zhang and Yao, Hailiang and others},
  booktitle={Forty-second International Conference on Machine Learning},
  year={2025}
}

@inproceedings{lian2025affectgpt,
  title={AffectGPT: A New Dataset, Model, and Benchmark for Emotion Understanding with Multimodal Large Language Models},
  author={Lian, Zheng and Chen, Haoyu and Chen, Lan and Sun, Haiyang and Sun, Licai and Ren, Yong and Cheng, Zebang and Liu, Bin and Liu, Rui and Peng, Xiaojiang and others},
  booktitle={Forty-second International Conference on Machine Learning},
  year={2025}
}

@article{wang2023incomplete,
  title={Incomplete multimodality-diffused emotion recognition},
  author={Wang, Yuanzhi and Li, Yong and Cui, Zhen},
  journal={Advances in Neural Information Processing Systems},
  volume={36},
  pages={17117--17128},
  year={2023}
}

@inproceedings{zhang2025moda,
  title={Moda: Modular duplex attention for multimodal perception, cognition, and emotion understanding},
  author={Zhang, Zhicheng and Xia, Wuyou and Zhao, Chenxi and Yan, Zhou and Liu, Xiaoqiang and Zhu, Yongjie and Qin, Wenyu and Wan, Pengfei and Zhang, Di and Yang, Jufeng},
  booktitle={Forty-second International Conference on Machine Learning},
  year={2025}
}

@inproceedings{zhang2025videmo,
  title={VidEmo: Affective-Tree Reasoning for Emotion-Centric Video Foundation Models},
  author={Zhang, Zhicheng and Wang, Weicheng and Zhu, Yongjie and Qin, Wenyu and Wan, Pengfei and ZHANG, Di and Yang, Jufeng},
  booktitle={Thirty-ninth Annual Conference on Neural Information Processing Systems},
  year={2025}
}

@inproceedings{li2023decoupled,
  title={Decoupled multimodal distilling for emotion recognition},
  author={Li, Yong and Wang, Yuanzhi and Cui, Zhen},
  booktitle={Proceedings of the IEEE/CVF Conference on Computer Vision and Pattern Recognition},
  pages={6631--6640},
  year={2023}
}

@inproceedings{poria2017context,
  title={Context-dependent sentiment analysis in user-generated videos},
  author={Poria, Soujanya and Cambria, Erik and Hazarika, Devamanyu and Majumder, Navonil and Zadeh, Amir and Morency, Louis-Philippe},
  booktitle={Proceedings of the 55th annual meeting of the association for computational linguistics (volume 1: Long papers)},
  pages={873--883},
  year={2017}
}

@inproceedings{tsai2019multimodal,
  title={Multimodal Transformer for Unaligned Multimodal Language Sequences},
  author={Tsai, Yao-Hung Hubert and Bai, Shaojie and Liang, Paul Pu and Kolter, J Zico and Morency, Louis-Philippe and Salakhutdinov, Ruslan},
  booktitle={Proceedings of the 57th Annual Meeting of the Association for Computational Linguistics},
  pages={6558--6569},
  year={2019}
}

@article{huang2026grading,
  title={Grading-inspired complementary enhancing for multimodal sentiment analysis},
  author={Huang, Zhijing and He, Wen-Jue and Hu, Baotian and Zhang, Zheng},
  journal={Information Fusion},
  pages={104174},
  year={2026},
  publisher={Elsevier}
}

@article{li2025uni,
  title={Uni-moe: Scaling unified multimodal llms with mixture of experts},
  author={Li, Yunxin and Jiang, Shenyuan and Hu, Baotian and Wang, Longyue and Zhong, Wanqi and Luo, Wenhan and Ma, Lin and Zhang, Min},
  journal={IEEE Transactions on Pattern Analysis and Machine Intelligence},
  volume={47},
  number={5},
  pages={3424--3439},
  year={2025},
  publisher={IEEE}
}

@inproceedings{li2023multi,
  title={A multi-modal context reasoning approach for conditional inference on joint textual and visual clues},
  author={Li, Yunxin and Hu, Baotian and Xinyu, Chen and Ding, Yuxin and Ma, Lin and Zhang, Min},
  booktitle={Proceedings of the 61st annual meeting of the association for computational linguistics (volume 1: long papers)},
  pages={10757--10770},
  year={2023}
}

@inproceedings{wu2018learning,
  title={Learning to extract coherent summary via deep reinforcement learning},
  author={Wu, Yuxiang and Hu, Baotian},
  booktitle={Proceedings of the AAAI conference on artificial intelligence},
  year={2018}
}

@article{liu2024contrastive,
  title={Contrastive learning based modality-invariant feature acquisition for robust multimodal emotion recognition with missing modalities},
  author={Liu, Rui and Zuo, Haolin and Lian, Zheng and Schuller, Bj{\"o}rn W and Li, Haizhou},
  journal={IEEE Transactions on Affective Computing},
  volume={15},
  number={4},
  pages={1856--1873},
  year={2024},
  publisher={IEEE}
}

@inproceedings{hu2025chain,
  title={Chain-talker: Chain understanding and rendering for empathetic conversational speech synthesis},
  author={Hu, Yifan and Liu, Rui and Ren, Yi and Yin, Xiang and Li, Haizhou},
  booktitle={Findings of the Association for Computational Linguistics: ACL 2025},
  pages={1988--2003},
  year={2025}
}

@inproceedings{liu2024emotion,
  title={Emotion rendering for conversational speech synthesis with heterogeneous graph-based context modeling},
  author={Liu, Rui and Hu, Yifan and Ren, Yi and Yin, Xiang and Li, Haizhou},
  booktitle={Proceedings of the AAAI Conference on Artificial Intelligence},
  pages={18698--18706},
  year={2024}
}

@inproceedings{liu2025hardness,
  title={Hardness-aware dynamic curriculum learning for robust multimodal emotion recognition with missing modalities},
  author={Liu, Rui and Zuo, Haolin and Lian, Zheng and Yuan, Hongyu and Fan, Qi},
  booktitle={Proceedings of the 33rd ACM International Conference on Multimedia},
  pages={5755--5764},
  year={2025}
}

\clearpage
\appendix

\section*{Appendix}  
\addcontentsline{toc}{chapter}{Appendix Contents} 
\etocsettocstyle{}{}  
\localtableofcontents 

\newpage

\section{Limitations and Future Work}
\label{appendix:limitations}
While this paper represents the first work exploring the role of reinforcement learning in open-vocabulary emotions, several limitations remain, which we leave to future work. First, we observe that adding a thinking process leads to performance degradation. Future work will involve designing high-quality cold-start data and conducting an in-depth analysis to determine which samples or emotional states necessitate a thinking process. Second, reinforcement learning yields performance gains for basic and fine-grained emotions but leads to slight degradation in sentiment analysis. Future work will focus on balancing reinforcement learning performance across different tasks. Additionally, we observe that increasing the amount of training data in reinforcement learning leads to progressive performance gains. Future work will focus on collecting more data for reinforcement learning to establish the upper-bound performance of our proposed method.


\section{Reproducibility Statement}
\label{appendix:reproducibility_statement}
In this paper, we have made every effort to ensure the reproducibility of our work. Our code is available at: \href{https://github.com/zeroQiaoba/AffectGPT/tree/master/AffectGPT-RL}{https://github.com/zeroQiaoba/AffectGPT/tree/master/AffectGPT-RL}.


\section{Ethics Statement}
\label{appendix:ethical_statement}
This paper does not involve the collection of new data or the hiring of annotators. We utilize MER-Caption+ \cite{lian2025affectgpt}, MER2025-OV \cite{lian2025mer}, OV-MERD+ \cite{lian2025affectgpt}, and MER-UniBench \cite{lian2025affectgpt}, with permission from the dataset owners. Furthermore, all datasets are employed in accordance with their licenses. Therefore, this paper raises no ethical concerns.

\section{Social Impact}
\label{appendix:social_impact}
Emotion plays an important role in communication, conveying human intentions and thoughts. OV-MER aims to capture subtle differences in emotion representation, serving as an important complement to existing representation methods. This paper centers on OV-MER and reveals the impact of reinforcement learning on OV-MER. Research surrounding OV-MER promises significant future social impact, particularly in enhancing the human-computer interaction experience.

\section{Novelty Statement}
\label{appendix:novelty}
There are currently two main types of emotion tasks: \emph{discriminative emotions} and \emph{generative emotions}. The former primarily involves tasks with a constrained label space, such as basic or dimensional emotions \cite{fang2025catch,sun2024svfap,wang2023incomplete}. The latter imposes no restrictions on the label space, giving rise to the emerging area of open-vocabulary emotions \cite{lian2025ov}. Recent research has shown a growing interest in shifting from \emph{discriminative emotions} to \emph{generative emotions}, driven by the need for more nuanced emotional representations \cite{cheng2024emotion,lian2025affectgpt,zhang2025videmo}. However, effectively addressing \emph{generative emotions} remains an open challenge. This paper focuses on open-vocabulary emotions, a specific subtask within \emph{generative emotions}, and proposes leveraging reinforcement learning as a solution. During our investigation, several practical issues emerged, including the design of reward functions and strategies to mitigate reward hacking. By tackling these challenges, our experimental results demonstrate the effectiveness of reinforcement learning in handling \emph{generative emotions}. Therefore, this paper should not be viewed merely as an application of reinforcement learning, but rather as a significant milestone in advancing \emph{generative emotions} research. This work will inspire more researchers to consider reinforcement learning as a foundational tool for addressing \emph{generative emotions}.


\section{EW-based Metric}
\label{appendix:ew_metric}

\subsection{Emotion Wheels}
\label{appendix:ew}
\begin{figure}[h]
    \begin{center}
        \begin{subcaptionbox}{W1}[0.26\linewidth]
            {\includegraphics[width=\linewidth]{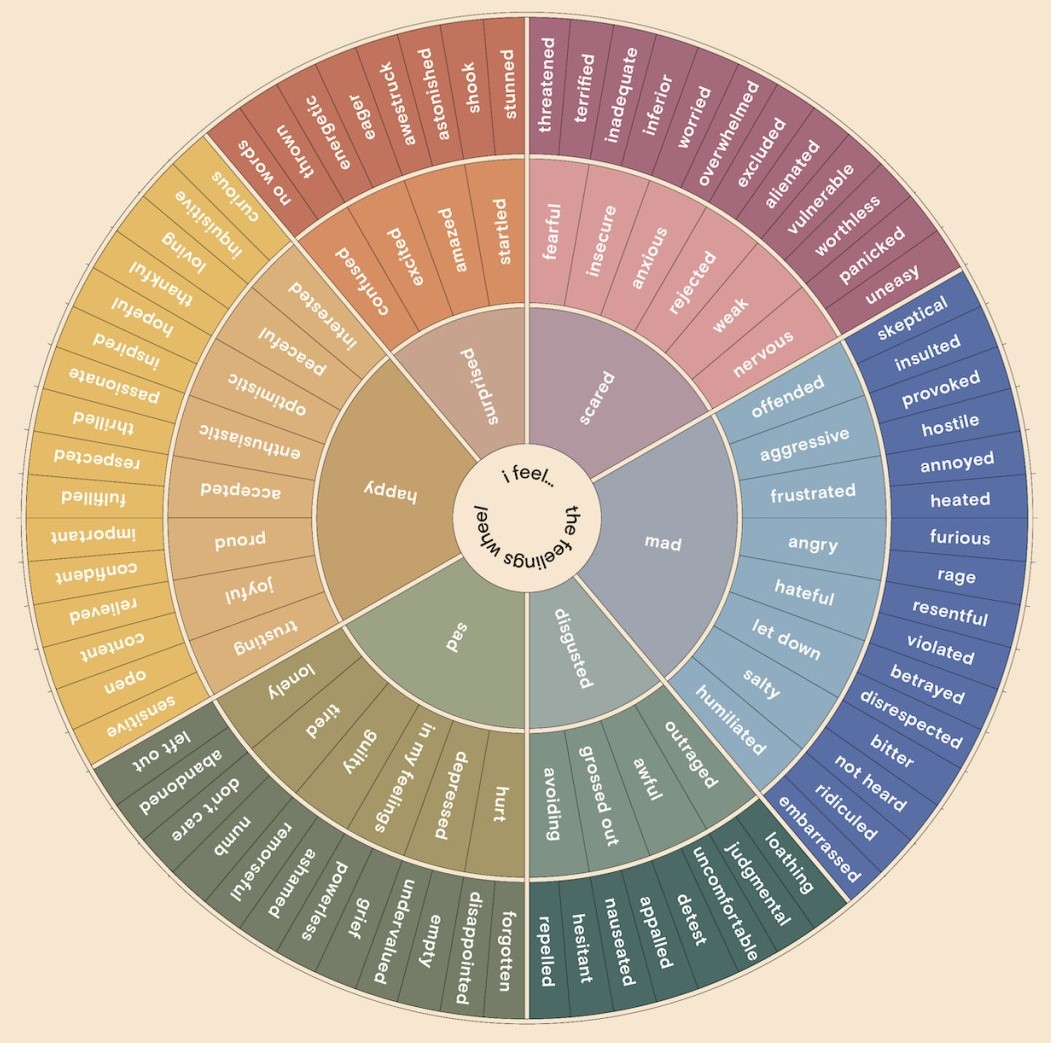}}
        \end{subcaptionbox}
        \begin{subcaptionbox}{W2}[0.26\linewidth]
            {\includegraphics[width=\linewidth]{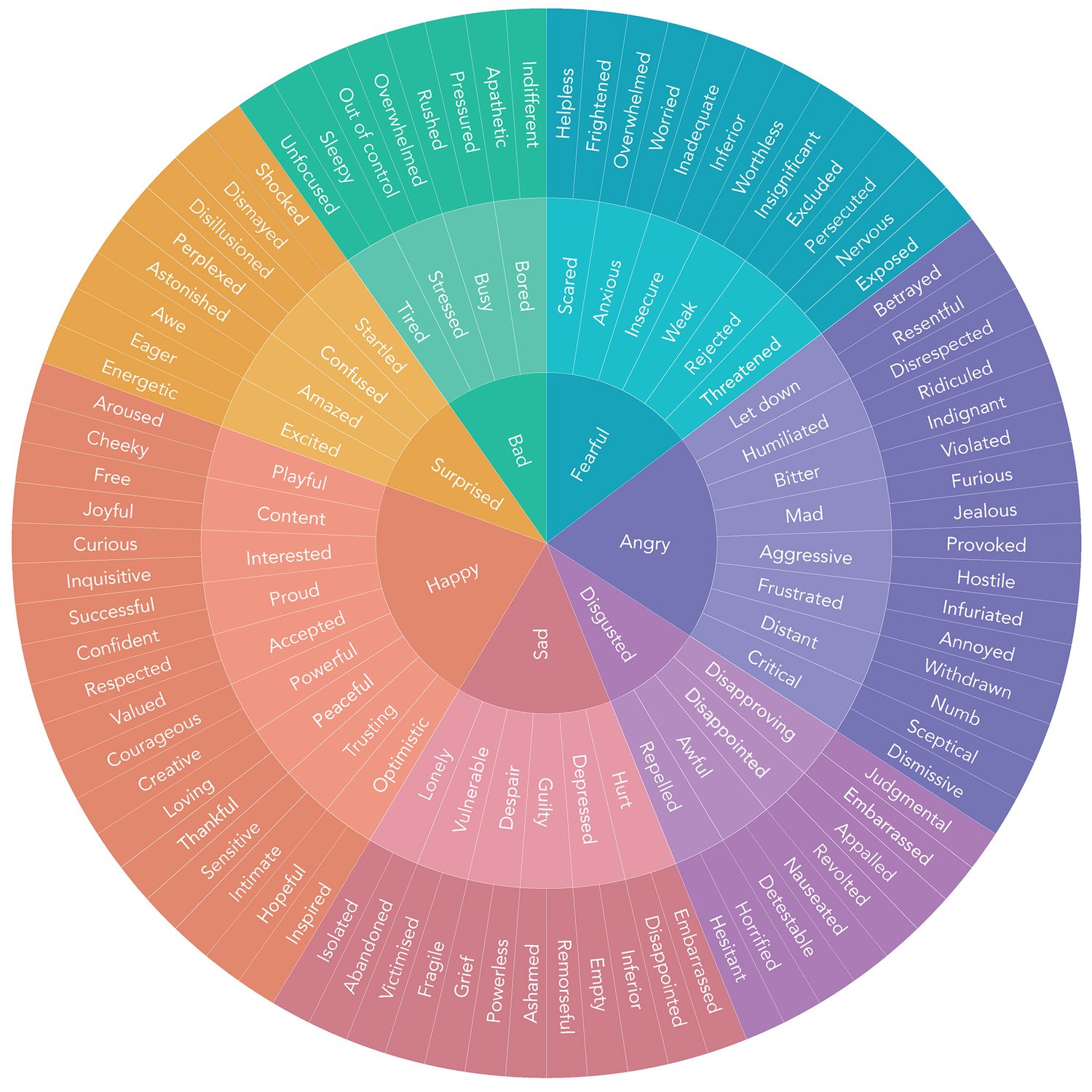}}
        \end{subcaptionbox}
        \begin{subcaptionbox}{W3}[0.26\linewidth]
            {\includegraphics[width=\linewidth]{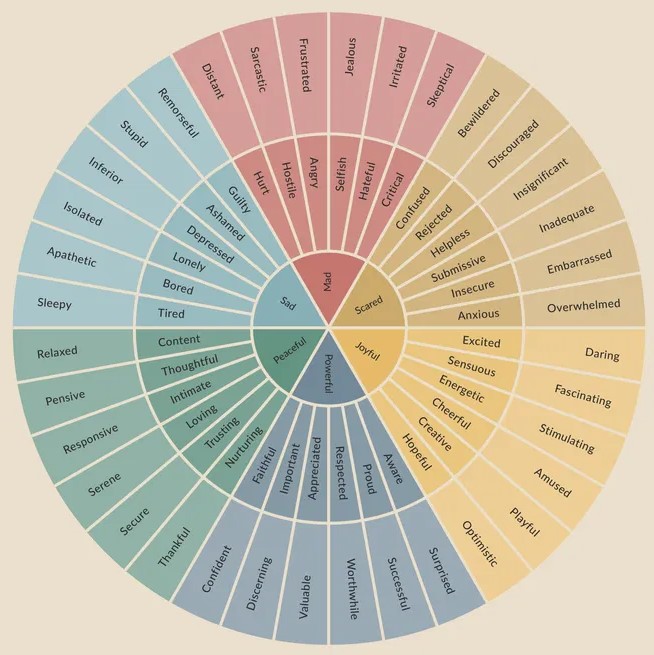}}
        \end{subcaptionbox}
        \begin{subcaptionbox}{W4}[0.26\linewidth]
            {\includegraphics[width=\linewidth]{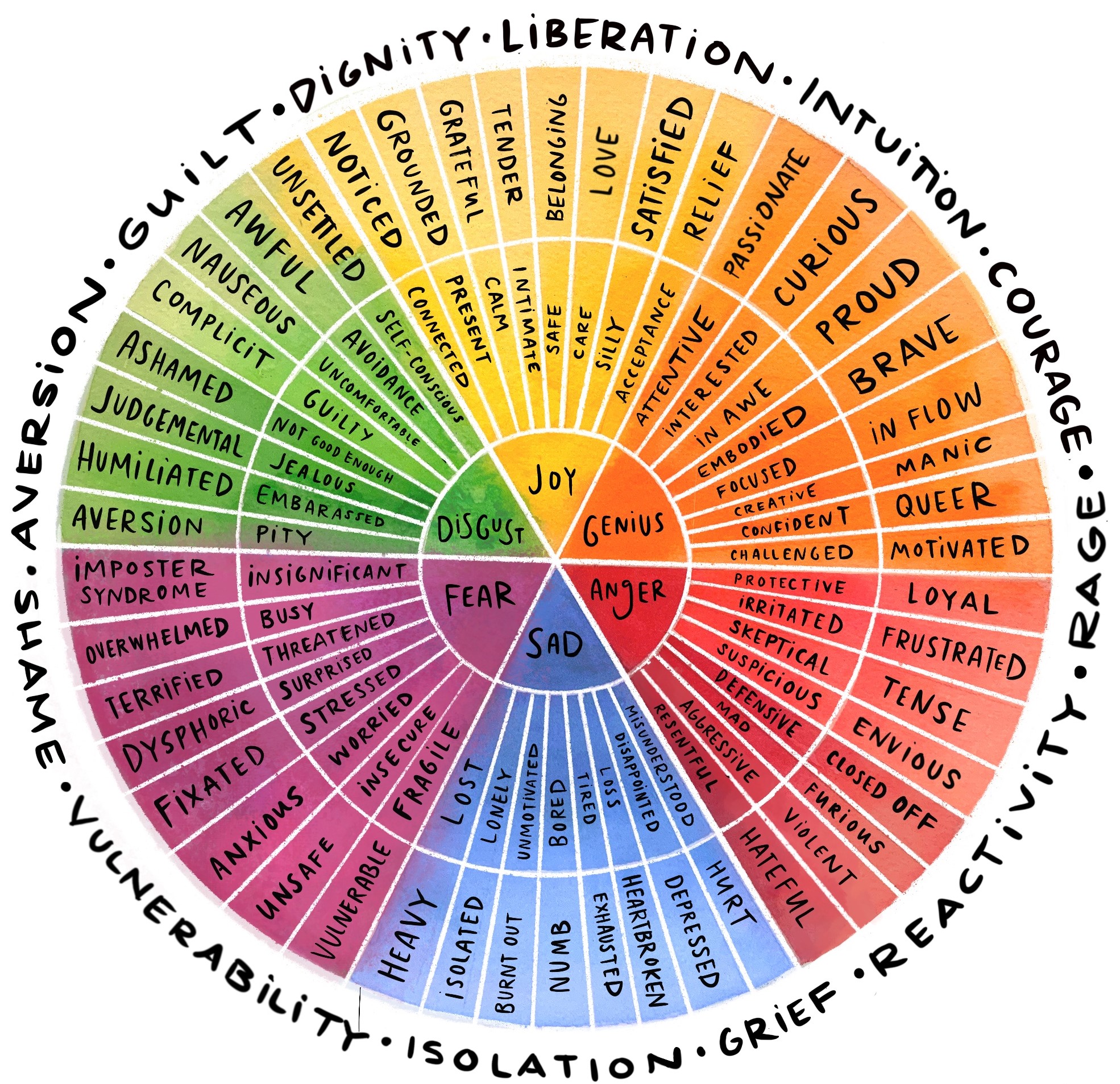}}
        \end{subcaptionbox}
        \begin{subcaptionbox}{W5}[0.26\linewidth]
            {\includegraphics[width=\linewidth]{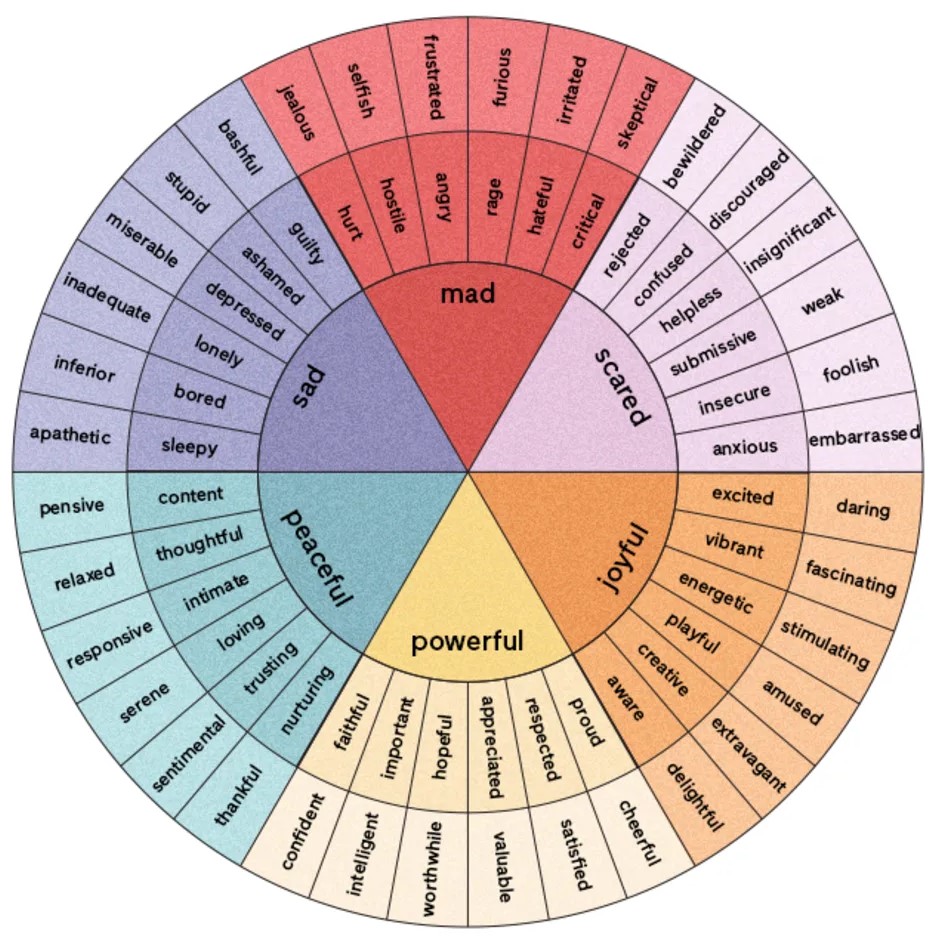}}
        \end{subcaptionbox}
    \end{center}
    \caption{\textbf{Emotion wheels}. This paper uses five emotion wheels to calculate EW-based metrics, ensuring a fair comparison with previous work \cite{lian2025affectgpt,lian2025ov}.}
    \label{fig:ew}
\end{figure}

The emotion wheel is a structured visual tool designed to categorize and identify human emotions. It organizes emotions into core categories (the inner sections of the wheel) and their nuanced variations (the outer sections), providing intuitive insights into emotional structure. This paper uses five emotion wheels to compute EW-based metrics, aligning with prior work \cite{lian2025affectgpt} to ensure fair comparisons. Figure \ref{fig:ew} illustrates these emotion wheels.

\subsection{Calculation Formula}
\label{appendix:ew_calculate}
We present the calculation formula for EW-based metrics \cite{lian2025ov}. The computation method primarily consists of two components: eliminating the influence of synonyms and defining set-level metrics.

\paragraph{Handling Synonyms.}
To mitigate the impact of synonyms, we employ a three-level hierarchical grouping strategy:

\textbf{(L1)} We map different forms of words to their base form. For example, \emph{happier} and \emph{happiness} are mapped to \emph{happy}. To achieve this, for a base word $w$, we first use LLM to generate various forms of $w$, with the output denoted as $l_1(w)$. Since different results may be produced across different calls, we repeat this process ten times and merge all the generated words, resulting in $\hat{l}_1(w)$. To map them back to their base form, we simply map $\hat{l}_1(w)$ to $w$, and we refer to this mapping function as $F_{l_1}(\cdot)$.
	
\textbf{(L2)} We map synonyms to a unified word. For example, \emph{happy} and \emph{joyful} are mapped to \emph{happy}. To achieve this, we use LLM to generate synonyms for a given word $w$, repeat the process multiple times, and then merge all the generated words, obtaining $\hat{l}_2(w)$. To map them to their unified form, we map each element in $\hat{l}_2(w)$ to $w$, and we refer to this mapping function as $F_{l_2}(\cdot)$.
	
\textbf{(L3)} The emotion wheel offers natural grouping information for emotions \cite{plutchik1980general}. Since there is no consensus on the emotion wheel, we adopt $K=5$ distinct emotion wheels, consistent with prior work \cite{lian2025affectgpt,lian2025ov} to ensure fair comparisons. Figure~\ref{fig:ew} visualizes all five emotion wheels. For each wheel $w_k$, we map all outer labels to their corresponding inner labels. This function is denoted as $F_{l_3}^{w_k}(\cdot)$.

Finally, the clustering functions can be summarized as:
\begin{equation}
G_{w_k}(\cdot) = F_{l_3}^{w_k}{\left(F_{l_2}\left(F_{l_1}\left(\cdot\right)\right)\right)}, k \in [1, K].
\end{equation}

\paragraph{Set-level Metrics.} 
Since we do not restrict the number of predicted and annotated emotions, we employ set-level evaluation metrics. Specifically, suppose the dataset contains $N$ samples. For a sample $x_i$, the true labels are $\mathbf{Y}_i=\{y_i^j\}_{j=1}^{n_i}$ and the predicted labels are $\mathbf{\hat{Y}}_i=\{\hat{y}_i^j\}_{j=1}^{\hat{n}_i}$. Here, $n_i$ and $\hat{n}_i$ are the number of predicted and annotated emotions, respectively. The evaluation metrics are defined as follows:
\begin{equation}
\mbox{Precision}_{\mbox{s}}^{k} = \frac{1}{N}\sum_{i=1}^{N}\frac{\left|G_{w_k}( \mathbf{Y}_i ) \cap G_{w_k}(\mathbf{\hat{Y}}_i)\right|}{\left|G_{w_k}(\mathbf{\hat{Y}}_i)\right|},
\end{equation}
\begin{equation}
\mbox{Recall}_{\mbox{s}}^{k} = \frac{1}{N}\sum_{i=1}^{N}\frac{\left|G_{w_k}( \mathbf{Y}_i ) \cap G_{w_k}(\mathbf{\hat{Y}}_i)\right|}{\left|G_{w_k}(\mathbf{{Y}}_i)\right|},
\end{equation}
\begin{equation}
\mbox{F}_{\mbox{s}}^{k} = 2\times\frac{\mbox{Precision}_{\mbox{s}}^{k}\times\mbox{Recall}_{\mbox{s}}^{k}}{\mbox{Precision}_{\mbox{s}}^{k}+\mbox{Recall}_{\mbox{s}}^{k}}.
\end{equation}

\textbf{In this set-wise operation, we automatically remove duplicate emotion words.}
Finally, we compute the average F1-score across all emotion wheels as the final score:
\begin{equation}
\text{EW}\left(\mathbf{Y}_i, \mathbf{\hat{Y}}_i\right) = \frac{1}{K}\sum_{k=1}^{K}\mbox{F}_{\mbox{s}}^{k}.
\end{equation}

\subsection{Non-Differentiability Analysis}
\label{appendix:ew_nondiff}
The non‑differentiability of EW‑based metrics stems from three compounding factors.
First, the predicted label set $\hat{\mathbf{Y}}_i$ is obtained via stochastic sampling from the model’s logits, which is inherently non-differentiable. Second, the three‑level clustering functions $F_{l_1}(\cdot)$, $F_{l_2}(\cdot)$, and $F_{l_3}(\cdot)$ are discrete word‑to‑word lookup mappings, where gradients could not propagate through these discrete clustering functions. Third, set‑level evaluation depends on cardinality operations. For instance, the intersection size $|G_{w_k}(\mathbf{Y}_i)\cap G_{w_k}(\hat{\mathbf{Y}}_i)|$ and the prediction-set size $|G_{w_k}(\hat{\mathbf{Y}}_i)|$ are integer‑valued functions of discrete token identities, yielding zero gradient with respect to continuous model parameters. Collectively, these three obstacles render EW‑based metrics non‑differentiable, precluding their direct optimization via gradient backpropagation and motivating our adoption of reinforcement learning.

\section{Details of Alignment Reward}
\label{appendix:alignment}

\paragraph{Emotion Word Extraction.}
To compute the alignment reward, we need to extract emotion words $e^t$ from $o^t$. To do so, we input $o^t$ into Qwen2.5-7B using the following prompt: \textcolor[rgb]{0.93,0.0,0.47}{\emph{Please assume the role of an expert in the field of emotions. We provide clues related to the emotional state of a character. Based on the provided clues, please identify the character's emotional states. Separate different emotional categories with commas, and output only the clearly identifiable emotional categories in a list format. If no emotions can be identified, please return an empty list.}} We select Qwen2.5-7B as the LLM backbone to ensure consistency with baselines \cite{lian2025affectgpt,lian2025ov}. This choice mitigates potential biases arising from the use of different LLMs, thereby ensuring a fair comparison with baselines.

\paragraph{Similarity Calculation.}
Alignment reward involves computing the similarity between two emotion sets. In practice, we use the \emph{Jaccard Similarity Coefficient}, which measures the similarity between two sets by comparing the size of their intersection to the size of their union. For clarity, we adopt the notation defined in Appendix \ref{appendix:ew_metric}. The corresponding calculation formula is provided below:
\begin{equation}
\mbox{Similarity}_{\mbox{s}}^{k} = \frac{1}{N}\sum_{i=1}^{N}\frac{\left|G_{w_k}( \mathbf{Y}_i ) \cap G_{w_k}(\mathbf{\hat{Y}}_i)\right|}{\left|G_{w_k}( \mathbf{Y}_i ) \cup G_{w_k}(\mathbf{\hat{Y}}_i)\right|},
\end{equation}
\begin{equation}
    \mathrm{is\_similar}\left(\mathbf{Y}_i, \mathbf{\hat{Y}}_i\right)=\frac{1}{K}\sum_{k=1}^{K}\mbox{Similarity}_{\mbox{s}}^{k}.
\end{equation}

\section{Details of Perception Reward}
\label{appendix:perception}

\paragraph{Preference of Two Descriptions.}
In our implementation, we adopt Qwen2.5-Omni as the MLLM, owing to its superior performance in emotion preference decoding \cite{lian2026emoprefer}. The prompt used is as follows: \textcolor[rgb]{0.93,0.0,0.47}{\emph{We provide two descriptions for a given input. Please determine which one is better aligned with the input content. If both descriptions are equally aligned with the input content, please output ``tie''. Directly output the answer without additional reasoning.}}

\paragraph{Ranking of Multiple Descriptions.}
Suppose there are $N$ descriptions $\{d_i\}_{i=1}^N$, where each description $d_i$ is associated with a positive parameter $\theta_i$ quantifying its relative advantage. The probabilities that description $d_i$ wins, loses, or ties against description $d_j$ are defined as follows:
\begin{equation}
    P(d_i>d_j) = \frac{\theta_i}{\theta_i+\theta_j+\beta},\;P(d_i<d_j) = \frac{\theta_j}{\theta_i+\theta_j+\beta},\;P(d_i \sim d_j) = \frac{\beta}{\theta_i+\theta_j+\beta},
\end{equation}
where $\beta$ controls the likelihood of ties. The parameters $\theta$ can then be estimated by maximizing the log-likelihood function:
\begin{equation}
    \max \mathcal{L}(\theta) = \prod_{i<j}\left(\frac{\theta_i}{\theta_i + \theta_j+\beta}\right)^{\mathbf{\text{win}}_{ij}}\left(\frac{\theta_j}{\theta_i + \theta_j+\beta}\right)^{\text{lose}_{ij}}\left(\frac{\beta}{\theta_i + \theta_j+\beta}\right)^{\text{tie}_{ij}},
\end{equation}
where $\text{win}_{ij}$, $\text{lose}_{ij}$, and $\text{tie}_{ij}$ are indicator variables denoting the preference between $d_i$ and $d_j$, determined by the MLLM in the preceding paragraph. Finally, we rank the descriptions based on their relative strengths $\theta_i$. A higher value indicates better alignment with the input content.

\section{Baseline Details}
\label{appendix:baseline}
\paragraph{SECap \cite{xu2024secap}.}
This is a speech emotion captioning framework designed to describe speech emotions using natural language. Specifically, it employs HuBERT as the audio encoder to extract speech representations and then uses the Bridge-Net to obtain emotion-relevant speech features. Finally, these features are fed into LLMs, which generate natural language captions describing speech emotions.

\paragraph{SALMONN \cite{tang2023salmonn}.}
This model is capable of processing both speech and audio. Specifically, it employs Whisper \cite{radford2023robust} to encode speech and BEATs \cite{chen2023beats} to encode audio. The resulting features are then integrated into the LLM via a window-level Q-Former. Additionally, a LoRA adapter is applied to enhance the LLM’s instruction-following and question-answering capabilities.

\paragraph{Qwen-Audio \cite{chu2023qwen}.}
This model is capable of handling 30 different tasks and a variety of audio types, demonstrating universal audio understanding capabilities. Specifically, it employs an audio encoder to map input audio into hidden features. To address interference between different tasks, it adopts a multi-task input format. The output features are then passed to LLMs for answer generation.

\paragraph{Otter \cite{li2025otter}.}
This model is built upon OpenFlamingo \cite{awadalla2023openflamingo} and trained on the MIMIC-IT dataset \cite{li2025otter}, demonstrating strong capabilities in multimodal perception, reasoning, and in-context learning.

\paragraph{Chat-UniVi \cite{jin2024chat}.}
This is a unified vision-language model capable of processing both images and videos. It employs dynamic visual tokens that capture spatial details in images and temporal relationships in videos.

\paragraph{Video-LLaVA \cite{lin2024video}.}
It aligns images and videos before projection, enabling the LLM to learn from a unified visual representation and equipping it with the ability to comprehend images and videos.

\paragraph{Video-ChatGPT \cite{maaz2024video}.}
This model averages frame-level features across temporal and spatial dimensions, extracting both spatial and temporal video features. These features are then projected into the text token space, enabling the LLM to generate responses.

\paragraph{VideoChat \cite{li2025videochat}.}
It uses video foundation models to encode videos as embeddings, then employs LLMs for response generation. After instruction fine-tuning, the model demonstrates promising capabilities across various video applications.

\paragraph{VideoChat2 \cite{li2024mvbench}.}
This model leverages diverse instruction-tuning data and employs a three-stage progressive multimodal training approach, comprising vision-language alignment, vision-language connection, and instruction tuning. As a result, it outperforms leading models across multiple tasks.

\paragraph{LLaMA-VID \cite{li2024llama}.}
This model represents each image with two tokens: a context token and a content token. The context token encodes the global image information based on user input, while the content token captures visual details within the image. By condensing each image into just a few tokens, the model can efficiently understand and process long videos.

\paragraph{mPLUG-Owl \cite{ye2023mplug}.}
This approach equips LLMs with vision-language capabilities through a two-stage training paradigm, enabling strong performance on instruction-following, visual understanding, and reasoning tasks.

\paragraph{PandaGPT \cite{su2023pandagpt}.}
It uses ImageBind \cite{girdhar2023imagebind} to map multiple modalities into a shared embedding space, then employs a projector to transform these features into the LLM's input space, and finally leverages the LLM to generate responses.

\paragraph{OneLLM \cite{han2024onellm}.}
This model aligns eight modalities with language through a unified framework. Its architecture consists of a universal encoder, a universal projection module, and modality-specific tokens. Additionally, it employs a progressive multimodal alignment pipeline, enabling it to comprehend eight input types and generate responses based on user queries.

\paragraph{R1-Omni \cite{zhao2025r1}.}
This framework also leverages reinforcement learning for emotion recognition. Unlike AffectGPT-RL, which focuses on open-vocabulary emotions, this work targets discriminative emotions. Due to the different objectives, the reward functions used in R1-Omni differ from those employed in AffectGPT-RL. Appendix~\ref{appendix:novelty} provides a more detailed comparison.

\paragraph{Emotion-LLaMA \cite{cheng2024emotion}.}
This framework follows the architecture of traditional MLLMs and consists of three key components: encoders, projectors, and LLMs. Specifically, it inputs text, audio, peak frames, and three levels of visual cues (i.e., local, temporal, and global representations) and employs modality-specific encoders to convert these inputs into corresponding hidden features. These features are then aligned into a shared dimensional space with text embeddings through a linear projection mechanism. Finally, the LLM integrates information from all modalities for emotion recognition and reasoning. Experimental results on benchmark datasets demonstrate that this framework achieves improved performance over previous solutions.

\paragraph{AffectGPT \cite{lian2025affectgpt}.}
This framework differs from previous MLLMs, which typically delegate the entire multimodal fusion process to the language model. Instead, it relocates cross-modal interaction outside the language model and incorporates a pre-fusion operation to enhance multimodal integration, emphasizing the inherently multimodal nature of human emotions. Specifically, the framework introduces two types of pre-fusion mechanisms: Q-Former and attention mechanisms. Both modules are computationally efficient and introduce only a small number of trainable parameters. Experimental results demonstrate the effectiveness of AffectGPT in generalized emotion understanding, including sentiment analysis, basic emotion recognition, and fine-grained emotion detection.

\section{Dataset Details}
\label{appendix:dataset}

\subsection{Training Datasets}

\paragraph{MER-Caption+ \cite{lian2025affectgpt}.}
It is a large-scale descriptive emotion dataset comprising approximately 31K samples. To reduce annotation costs, it introduces a novel \emph{model-led human-assisted} strategy that leverages human priors to guide both description generation and sample filtering. Specifically, in the description generation stage, the pre-trained models are carefully selected to maximize performance on fine-grained emotion detection. During sample filtering, a two-level filtering process is employed to remove samples with mismatched audio and video, abnormally long descriptions, and those that perform poorly on sentiment analysis and basic emotion recognition.

\paragraph{MER2025-OV \cite{lian2025mer}.}
It is specifically designed for OV-MER. Unlike the automatically annotated MER-Caption+, MER2025-OV is a \emph{high-quality} dataset created through a purely manual annotation process. It employs a multi-round annotation strategy. Initially, a pool of candidate emotions is provided, and four experts in affective computing are asked to select the correct labels and add any missing ones. All selected labels are combined to form the candidate set for the next round. In the second round, four annotators perform another round of labeling, and only labels selected by at least two annotators are retained. This two-stage approach ensures both completeness, by accepting labels chosen by any annotator in the first round, and accuracy, by filtering for consensus in the second. Through this rigorous process, MER2025-OV guarantees the high quality of its annotated labels.

\subsection{Testing Datasets}

\paragraph{OV-MERD+ \cite{lian2025affectgpt}.}
OV-MERD+ is a dataset specifically designed for OV-MER, which is an extended version of OV-MERD \cite{lian2025ov}. The original OV-MERD dataset consists of 332 samples, all of which are fully annotated by humans for open-vocabulary emotions. Following a similar annotation process, OV-MERD+ expands the dataset to a total of 532 samples. In this study, we employ OV-MERD+ as the test set to evaluate the performance of different models on open-vocabulary emotion detection.

\paragraph{MER-UniBench \cite{lian2025affectgpt}.}
\label{appendix:merunibench}
MER-UniBench is a benchmark that encompasses three representative tasks: sentiment analysis, basic emotion recognition, and fine-grained emotion detection. Specifically, sentiment analysis focuses on binary classification into positive and negative sentiments. Basic emotion recognition restricts the label space to a predefined set of basic emotion categories. In contrast, fine-grained emotion detection removes these constraints, allowing predictions across any category. This benchmark is designed to assess generalized emotion understanding capabilities. Table~\ref{tab:merunibench} summarizes the data composition for each task.

\begin{table}[h]
	\centering
	\caption{\textbf{MER-UniBench.} This table presents the data components for sentiment analysis, basic emotion recognition, and fine-grained emotion detection. Totally, it includes 12,799 test samples.}
    \label{tab:merunibench}
    \resizebox{\linewidth}{!}{
        \begin{tabular}{l|c|r|l}
            \toprule
            \textbf{Raw Dataset} & \textbf{Selected Subset} & \textbf{Number of Samples} & \textbf{Emotion Space} \\
            \midrule
            \rowcolor{gray!20}
            \multicolumn{4}{c}{\emph{Sentiment Analysis}} \\
            \midrule
            CMU-MOSI \cite{zadeh2017tensor}      & Test & 686   & \textcolor[rgb]{0.93,0.0,0.47}{\emph{positive, negative}} \\
            CMU-MOSEI \cite{zadeh2018multimodal} & Test & 4,659 & \textcolor[rgb]{0.93,0.0,0.47}{\emph{positive, negative}} \\
            CH-SIMS \cite{yu2020ch}              & Test & 457   & \textcolor[rgb]{0.93,0.0,0.47}{\emph{positive, negative}} \\
            CH-SIMS v2 \cite{liu2022make}        & Test & 1,034 & \textcolor[rgb]{0.93,0.0,0.47}{\emph{positive, negative}} \\
            \midrule
            \rowcolor{gray!20}
            \multicolumn{4}{c}{\emph{Basic Emotion Recognition}} \\
            \midrule
            MER2023 \cite{lian2023mer}           & MER-MULTI & 411   & \textcolor[rgb]{0.93,0.0,0.47}{\emph{worry, happy, neutral, angry, surprised, sad}} \\
            MER2024 \cite{lian2024mer}           & MER-SEMI  & 1,169 & \textcolor[rgb]{0.93,0.0,0.47}{\emph{worry, happy, neutral, angry, surprised, sad}} \\
            IEMOCAP \cite{busso2008iemocap}      & Session5  & 1,241 & \textcolor[rgb]{0.93,0.0,0.47}{\emph{anger, happiness, sadness, neutral}} \\
            MELD \cite{poria2019meld}            & Test      & 2,610 & \textcolor[rgb]{0.93,0.0,0.47}{\emph{anger, joy, sadness, neutral, disgust, fear, surprise}} \\
            \midrule
            \rowcolor{gray!20}
            \multicolumn{4}{c}{\emph{Fine-grained Emotion Detection}} \\
            \midrule
            OV-MERD+ \cite{lian2025affectgpt}  & All & 532 & \textcolor[rgb]{0.93,0.0,0.47}{\emph{unrestricted label space}} \\
            \bottomrule
        \end{tabular}
    }
\end{table}

\subsection{Data Leakage Prevention}
\label{appendix:data_leakage}
To prevent data leakage, we perform a rigorous verification to ensure no overlap exists between the training and testing data. During this process, we identify 200 overlapping samples between MER2025-OV (training set) and OV-MERD+ (test set). Consequently, we remove these samples from MER2025-OV, resulting in 1,000 samples available for training (see Table \ref{tab:dataset}). Apart from this instance, we observe no other overlap between the training and testing sets. This procedure ensures that no data leakage exists between the datasets used for training and testing.

\subsection{Task Clarification}
\label{appendix:task}
This paper focuses on \emph{utterance-level} emotion recognition, as opposed to \emph{dialogue-level} emotion recognition. The key difference between these tasks lies in whether conversational history is taken into account. For \emph{utterance-level} emotion recognition, the input consists of a single video clip centered on one character \cite{tsai2019multimodal}. In contrast, \emph{dialogue-level} emotion recognition requires leveraging additional dialogue information \cite{poria2017context}. This distinction has led to the development of two types of benchmarks: the \emph{utterance-level} benchmark \cite{lian2026merbench} and the \emph{dialogue-level} benchmark \cite{poria2019emotion}.

In MER-UniBench, although certain datasets (e.g., MELD and IEMOCAP) contain dialogue information, others do not. Furthermore, even when dialogue context is available in datasets like MELD and IEMOCAP, its inclusion is not necessary and depends on the specific experimental setup \cite{tsai2019multimodal, poria2017context}. Consequently, MER-UniBench is designed as an \emph{utterance-level} benchmark, and this work adopts the same setting to ensure fair comparison with prior studies \cite{lian2025affectgpt}. Dialogue-level emotion recognition is beyond the scope of this paper and will be explored in the future.

\section{Role of RL across Different Architectures}
\label{appendix:impact_of_different_backbones}

This paper adopts Qwen2.5-7B-Instruct as the default LLM. In Table \ref{tab:impact_of_different_backbones}, we replace Qwen2.5-7B-Instruct with various alternative LLMs to investigate the role of RL across different architectures. Experimental results are presented in Table \ref{tab:impact_of_different_backbones}. The second column reports the SFT baseline results, while the third column shows the results after applying RL. These results demonstrate that reinforcement learning yields consistent performance gains across different model architectures, underscoring the effectiveness of RL training for tackling open-vocabulary emotion recognition.

\begin{table}[h]
	\centering
	\caption{\textbf{Role of RL across different architectures.} To evaluate the generalization of RL, we replace the default Qwen2.5-7B-Instruct backbone in AffectGPT-RL with various alternative LLMs. Performance is assessed on OV-MERD+ to quantify the impact on open-vocabulary emotion recognition.}
    \label{tab:impact_of_different_backbones}
        \begin{tabular}{l|c|c|c}
            \toprule
            \textbf{LLM Backbone} & \textbf{SFT} & \textbf{RL} & \textbf{$\Delta$ Gain} \\
            \midrule
            Qwen2.5-7B-Instruct      & 62.52 & 68.39 & \textcolor{green!60!black}{+5.87} \\
            Qwen2-7B-Instruct        & 59.63 & 67.02 & \textcolor{green!60!black}{+7.39} \\
            Baichuan2-7B-Chat        & 61.35 & 66.56 & \textcolor{green!60!black}{+5.21} \\
            Meta-Llama-3-8B-Instruct & 62.98 & 66.24 & \textcolor{green!60!black}{+3.26} \\
            \bottomrule
        \end{tabular}
\end{table}

\section{Impact of Training Data Scale in RL}
\label{appendix:more_rl_data}
In the main experiments, we use MER2025-OV for reinforcement learning (see Table~\ref{tab:dataset}). To investigate the impact of data scale, we additionally annotated 1,000 open-vocabulary emotion samples, termed OV-New, and combined them with MER2025-OV to construct an extended training set of 2,000 samples. The corresponding experimental results are presented in Table~\ref{tab:more_rl_data}. From this table, we observe that increasing the amount of RL data leads to limited performance gains, indicating that the performance of RL has reached its upper bound.

\begin{table}[h]
	\centering
	\caption{\textbf{Impact of training data scale in RL.} We investigate whether increasing the volume of RL training data yields continued performance improvements.}
    \label{tab:more_rl_data}
        \begin{tabular}{l|c|cccc}
            \toprule
            \textbf{Dataset} & \textbf{\# Samples} & \textbf{Performance on OV-MERD+} \\
            \midrule
            MER2025-OV   \cite{lian2025mer}  & 1,000  & 68.39 \\
            MER2025-OV + OV-New              & 2,000  & 68.91 \\
            \bottomrule
        \end{tabular}
\end{table}

\end{document}